\documentclass[runningheads,fleqn]{svmult}

\usepackage{makeidx}   % allows index generation
\usepackage{graphicx}  % standard LaTeX graphics tool
\usepackage{subeqnar}  % subnumbers individual equations
\usepackage{multicol}  % used for the two-column index
\usepackage{physmult}  % flushleft layout of captions etc.
\makeindex             % used for the subject index
\begin{document}
\title*{Quantum phase transitions from topology in momentum space}
\toctitle{Quantum phase transitions
\protect\newline  from topology in momentum space}
% allows explicit linebreak for the table of content
%
%
\titlerunning{Topology of quantum phase transitions}
% allows abbreviation of title, if the full title is too long
% to fit in the running head
%
\author{G.~E.~Volovik\inst{1,2}}
\institute{Low Temperature Laboratory, Helsinki University of Technology,
  P.O.Box 2200, FIN-02015
  HUT, Espoo, Finland
\and  Landau Institute for Theoretical Physics, Kosygina 2, 119334
  Moscow, Russia
}
\authorrunning{G.~E.~Volovik}
% if there are more than two authors,
% please abbreviate author list for running head
%
%

\maketitle              % typesets the title of the contribution

\vskip5mm
Many quantum condensed matter systems are strongly correlated and strongly 
interacting fermionic systems, which cannot be treated perturbatively.
However, physics which emerges in the low-energy corner does not depend on
the complicated details of the system and is relatively simple. It is
determined by the nodes in the fermionic spectrum, which are protected by
topology in momentum space (in some cases, in combination with the vacuum
symmetry). Close to the nodes the behavior of the system becomes
universal; and the universality classes are determined by the
toplogical invariants in momentum space. When one changes the parameters of
the system, the transitions are expected to occur between
the vacua with the same symmetry but which belong to different universality
classes. Different types of quantum phase transitions governed by topology
in momentum space are discussed in this  Chapter. They involve Fermi
surfaces, Fermi points, Fermi lines, and also the topological 
transitions between the fully gapped states. The consideration based on
the  momentum space topology of the Green's function is general and is
applicable to the vacua of relativistic quantum fields. This is
illustrated by the possible quantum phase transition governed by topology
of nodes in the spectrum of elementary particles of Standard Model.

\section{Introduction.} 

There are two schemes for the classification of
states in condensed matter physics and relativistic quantum fields:
classification by symmetry (GUT scheme) and  by momentum space topology 
(anti-GUT scheme). 

For the first classification method, a given state of the
system is characterized by a symmetry group $H$ which is a subgroup
of the symmetry group $G$ of the relevant physical laws. 
The thermodynamic phase transition
between equilibrium states is usually marked by a change of the
symmetry group $H$. This classification reflects the
phenomenon of spontaneously broken symmetry. In relativistic quantum
fields the chain of successive phase transitions, in which the large
symmetry group existing at high energy is reduced at low energy, 
is in the basis of the Grand
Unification models (GUT) \cite{UnificationModel,Unification}. In
condensed matter  the spontaneous symmetry breaking is a typical
phenomenon, and the thermodynamic states are also classified in terms of 
the subgroup $H$ of the relevant group $G$  (see e.g, the classification 
of superfluid and superconducting states  in Refs.
\cite{VolovikGorkov1985,VollhardtWoelfle}). The groups $G$ and $H$  are
also responsible for  topological defects,  which are determined by the
nontrivial elements of the homotopy groups $\pi_n(G/H)$; cf. Ref.
\cite{TopologyReview1}.

The second classification method reflects the opposite tendency -- 
the anti Grand
Unification (anti-GUT) -- when
instead of the symmetry breaking the symmetry gradually
emerges at low energy.  This
method deals with the ground states of the system at zero temperature
($T=0$), i.e., it is the classification of quantum vacua. The
universality classes of quantum vacua are determined by momentum-space
topology, which is also responsible for the type of the effective
theory, emergent physical laws and symmetries at low energy. Contrary to
the GUT scheme,  where the symmetry of the vacuum
state is primary giving rise to topology, in the anti-GUT
scheme the topology in the momentum space is primary while the vacuum
symmetry is the emergent phenomenon in the low energy corner.

At the moment, we live in the ultra-cold Universe. All the characteristic 
temperatures in our Universe are extremely small compared to the Planck
energy scale $E_{\rm P}$.
 That is why all the massive
fermions, whose natural mass must be of order  $E_{\rm P}$, are frozen 
out due to extremely small factor $\exp(-E_{\rm P}/T)$.  There is no
matter  in our Universe unless there are massless fermions, whose
masslessness is protected with extremely high accuracy. It is the
topology in the momentum space, which provides such protection.

For systems living in 3D space, there are four basic universality classes
of fermionic vacua provided by  topology in momentum space
\cite{VolovikBook,Horava}: 

(i) Vacua with fully-gapped fermionic excitations, such as semiconductors
and conventional superconductors.
 
(ii) Vacua with fermionic excitations
characterized by Fermi points -- points in 3D momentum space at which the
energy  of fermionic quasiparticle vanishes. Examples
are provided by superfluid $^3$He-A and also by the quantum vacuum of
Standard Model above the electroweak transition, where all elementary
particles are Weyl fermions with Fermi points in the spectrum. This
universality class manifests the phenomenon of emergent relativistic
quantum fields at low energy: close to the Fermi points the
fermionic quasiparticles behave as massless Weyl fermions, while the
collective modes of the vacuum interact with these fermions as gauge and
gravitational fields.

(iii) Vacua with fermionic excitations characterized by lines in  
3D momentum space or points in 2D momentum space. We call them Fermi 
lines, though in general it is better to characterize zeroes by
co-dimension,  which is the dimension of ${\bf p}$-space minus the
dimension of the manifold of zeros.  Lines in  
3D momentum space and points in 2D momentum space have  co-dimension 2: 
since
$3-1=2-0=2$; compare this with zeroes of class (ii) which have 
co-dimension
$3-0=3$. 
 The Fermi lines are
topologically stable only if some special symmetry is obeyed. Example is 
provided by the vacuum of the high
$T_c$ superconductors where the Cooper pairing into a $d$-wave state 
occurs. The nodal lines (or actually the point nodes in these effectively
2D systems) are stabilized by the combined effect of momentum-space
topology and time reversal symmetry.

(iv) Vacua with fermionic excitations characterized by Fermi surfaces. 
The representatives of this universality class are normal metals and
normal liquid $^3$He.  This universality class also manifests the
phenomenon of emergent physics, though non-relativistic: at low
temperature all the metals behave in a similar way, and this behavior is
determined by the Landau theory of Fermi liquid -- the effective theory
based on the existence of Fermi surface.  Fermi surface has co-dimension
1: in 3D system it is the surface (co-dimension $=3-2=1$), in 2D system
it is the line (co-dimension $=2-1=1$), and in 1D system it is the point
(co-dimension $=1-0=1$; in one dimensional system the Landau Fermi-liquid
theory does not work, but the Fermi surface survives). 

The possibility of the Fermi band class (v), where the energy vanishes in
the finite region of the 3D momentum space and thus zeroes have
co-dimension 0, has been also discussed
\cite{Khodel1990,NewClass,Shaginyan,Khodel2005}. It is believed that this 
the so-called Fermi condensate may occur in strongly interacting 
electron systems PuCoGA$_5$ and CeCoIn$_5$ \cite{Khodel2005b}.  
Topologically stable flat band may exist in the spectrum of fermion zero modes, i.e.
for fermions localized in the core of the topological objects \cite{Volovik1994}.

\begin{figure}[t]
\centerline{\includegraphics[width=0.90\linewidth]{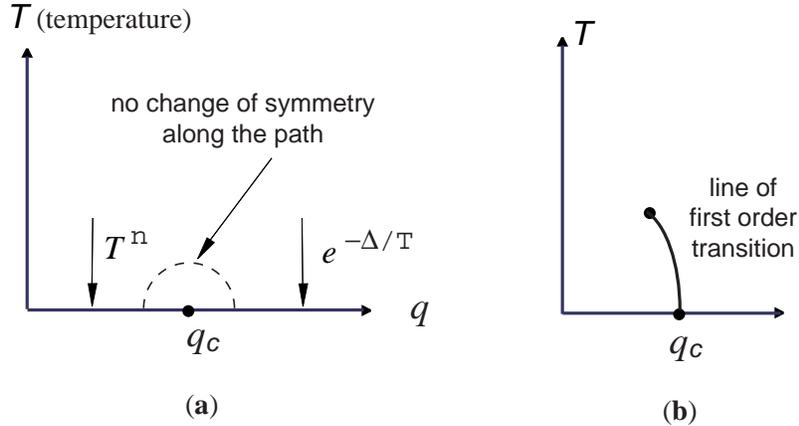}}
%\centerline{\epsfxsize=0.70\textwidth\epsfbox{QPT.eps}}
\medskip
\caption{Quantum phase transition between two ground states with the  same
symmetry but of different universality class -- gapless at $q<q_c$  and
fully gapped at $q>q_c$ -- as isolated point   ({\bf a})  as
the termination point of first order transition  ({\bf b})
 }
\label{QPTFig}
\end{figure}

The phase transitions which follow from this classification scheme are
quantum phase transitions which occur at $T=0$ \cite{Sachdev}. It may
happen that by changing some parameter
$q$ of the system we transfer the vacuum state from one universality
class  to another, or to the vacuum of the same universality class
but different topological quantum number, without changing its symmetry
group
$H$. The point
$q_c$, where this zero-temperature transition occurs, marks the quantum
phase transition. For $T\neq 0$, the second order phase transition is 
absent, as the two states belong to the same symmetry class $H$, but the
first order phase transition is not excluded. Hence, there is an isolated
singular point
$(q_c,0)$ in the
$(q,T)$ plane (Fig. \ref{QPTFig}({\bf a})), or the end point of the 
first order transition (Fig. \ref{QPTFig}({\bf b})). 

The quantum phase transitions which occur in classes 
 (iv) and (i) or between these classes are well known. 
In the class (iv) the corresponding quantum phase transition is known as
Lifshitz transition \cite{Lifshitz}, at which the Fermi surface changes
its topology or emerges from the fully gapped state of class (i),   see
Sec. \ref{LifshitzTransition}. The transition between the fully gapped
states   characterized by  different topological charges occurs in 2D
systems exhibiting the quantum Hall and spin-Hall effect: this is the
plateau-plateau transition between the states with different values of
the Hall (or spin-Hall) conductance (see Sec.
\ref{PlateauTransitions}).  The less known transitions involve 
nodes of co-dimension 3 
\cite{Book1,QPT,SplittingPreprint,Gurarie,BotelhoPWavw} (Sec.
\ref{FermiPoints} on Fermi points) and nodes of co-dimension 2
\cite{Botelho,Borkowski,Duncan,WenZee}   ( Sec.
\ref{FermiLines}  on nodal lines).
The quantum phase transitions involving the flat bands of class (v) 
are discussed in  Ref. \cite{Volovik1994}.

\section{Fermi surface and Lifshitz transition} 
\subsection{Fermi surface as a vortex in ${\bf p}$-space} 

\begin{figure}[t]
\centerline{\includegraphics[width=\linewidth]{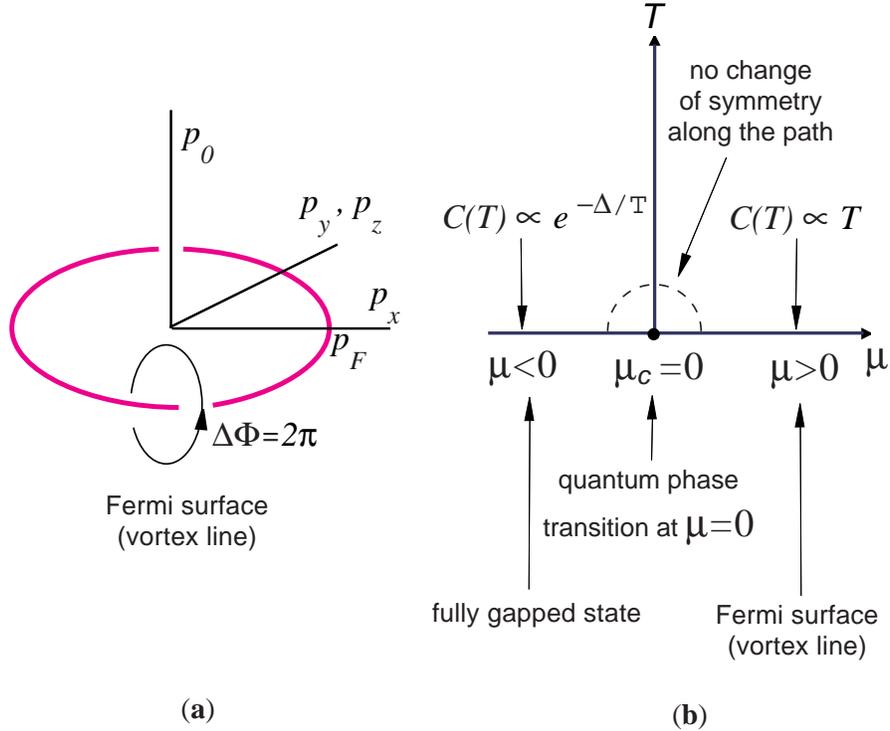}}
%\centerline{\epsfxsize=0.70\textwidth\epsfbox{FermiSurfQPT.eps}}
\medskip
\caption{({\bf a}) Fermi surface is a topological object in momentum space
--  a vortex loop. ({\bf b}) When the chemical potential $\mu$ decreases
the loop shrinks and disappears  at
$\mu<0$. The point $\mu=T=0$ marks the Lifshitz transition between the
gapless ground state at $\mu>0$ to the fully gapped vacuum at $\mu<0$}
\label{FermiSurfaceQPTFig}
\end{figure}

In ideal Fermi gases, the Fermi surface at $p=p_F=\sqrt{2\mu m}$ is the
boundary in ${\bf p}$-space between the occupied states ($n_{\bf p}=1$) 
at $p^2/2m <\mu$  and empty states ($n_{\bf p}=0$) at $p^2/2m
>\mu$. At this boundary (the surface in 3D momentum space) the energy  is
zero. What happens when the interaction between particles is introduced? 
Due to interaction the  distribution function
$n_{\bf p}$ of particles in the ground state is no longer exactly  1 or
0. However, it appears that the Fermi surface 
survives as the singularity in $n_{\bf p}$.  Such
stability of the Fermi surface comes from a topological property of the
one-particle Green's function at imaginary frequency:
\begin{equation}
G^{-1}=  i\omega -\frac{p^2}{2m} +\mu~.
\label{Propagator2}
\end{equation}
Let us for simplicity
skip one spatial dimension $p_z$ so that the Fermi surface becomes  the
line in 2D momentum space $(p_x,p_y)$; this does not change the
co-dimension of zeroes which remains  $1=3-2=2-1$. The Green's function
has singularities lying on a closed line
$\omega=0$,
$p_x^2+p_y^2=p_F^2$ in the 3D  momentum-frequency space 
$(\omega,p_x,p_y)$  (Fig. \ref{FermiSurfaceQPTFig}({\bf a})). This is
the line of the quantized vortex  in the momemtum space, since the phase
$\Phi$ of the Green's function
$G=|G|e^{i\Phi}$ changes by $2\pi N_1$ around the path embracing  any
element of this vortex line. In the considered case the phase winding
number  is
$N_1=1$.  If we add the third momentum dimension $p_z$ the vortex line
becomes the surface in the 4D momentum-frequency space
$(\omega,p_x,p_y,p_z)$  -- the Fermi surface -- but again the phase
changes by $2\pi$ along any closed loop empracing the element of the 2D
surface in the 4D  momentum-frequency space.

The winding number cannot change by continuous deformation of the 
Green's function:  the momentum-space vortex is robust toward any
perturbation. Thus the singularity of the Green's function on the Fermi
surface is preserved, even when interaction between fermions is
introduced. The invariant is the same for any space dimension, since the
co-dimension remains 1.

The Green function is generally a matrix with spin
indices. In addition, it may have the band indices (in the case of 
electrons in the periodic potential of crystals). In such a case the
phase of the Green's function becomes meaningless; however, the
topological property of the Green's function remains robust.  The general
analysis \cite{Horava} demonstrates that topologically stable Fermi
surfaces are described by the group $Z$ of integers. The  winding number
$N_1$ is expressed analytically  in terms of the Green's function
\cite{VolovikBook}:
\begin{equation}
N_1={\bf tr}~\oint_C {dl\over 2\pi i}  G(\mu,{\bf p})\partial_l
G^{-1}(\mu,{\bf p})~.
\label{InvariantForFS}
\end{equation}
Here the integral is taken over an arbitrary contour $C$ around  the
momentum-space vortex, and
${\bf tr}$ is the trace over the spin, band and/or other indices.

\subsection{Lifshitz transitions} 
\label{LifshitzTransition}

There are two scenarios of how to destroy the vortex loop in momentum 
space:  perturbative and non-perturbative. The non-perturbative 
mechanism of destruction  of the Fermi surface occurs for example  at the
superconducting transition, at which the  spectrum changes drastically
and  the gap appears. We shall consider this later  in Sec.
\ref{Superconductivity}, and  now let us concentrate on the perturbative
processes.

\subsubsection{Contraction and expansion of vortex loop in ${\bf p}$-space}

The Fermi surface cannot be destroyed by small perturbations,  since it is
protected by topology and thus is robust to perturbations.  But the Fermi
surface can be removed by large perturbations  in the processes which
reproduces the processes occurring for  the real-space counterpart of the
Fermi surface -- the loop of quantized vortex in superfluids and
superconductors. The vortex ring can continuously shrink  to a point and
then disappear, or continuously expand and leave the momentum space.  
The first scenario occurs when one continuously  changes
the chemical potential from the positive to the negative value: at
$\mu<0$ there is no vortex loop in momentum space and the ground state
(vacuum) is fully gapped.  The point $\mu=0$ marks the quantum phase
transition -- the Lifshitz transition -- at which the topology of the
energy spectrum changes  (Fig. \ref{FermiSurfaceQPTFig}({\bf b})). At this
transition the symmetry of the ground state does not changes. The second
scenario of the quantum phase transition to the fully gapped states occurs
when  the inverse mass $1/m$
 in Eq.(\ref{Propagator2}) crosses zero.

\begin{figure}[t]
\centerline{\includegraphics[width=\linewidth]{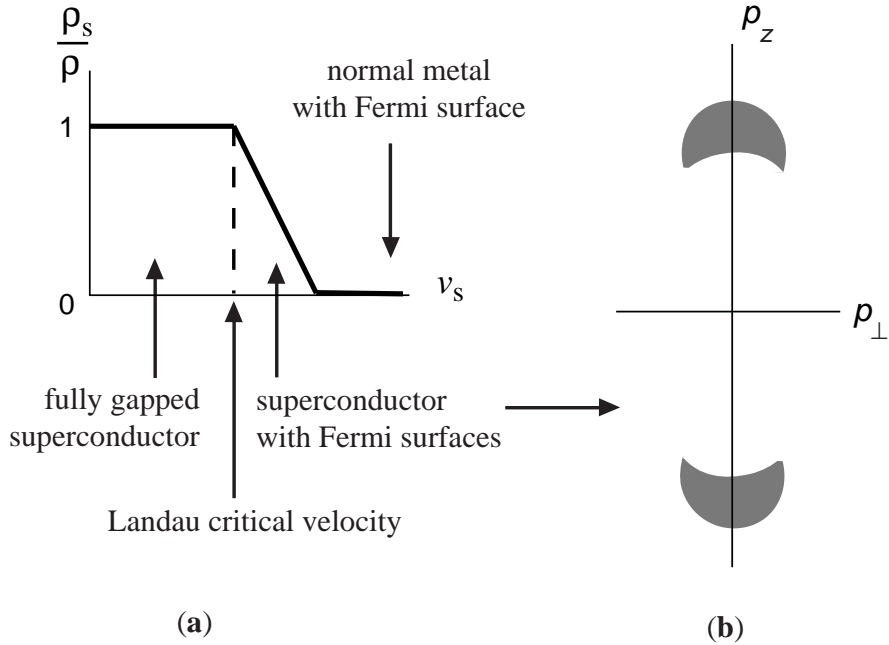}}
%\centerline{\epsfxsize=0.70\textwidth\epsfbox{CritVel.eps}}
\medskip
\caption{Illustration of Lifshitz transition in superfluid/superconductor
at Landau critical velocity.  ({\bf a}) In the presence of the superfluid
motion with velocity ${\bf v}_s$, the spectrum of quasiparticles  is
Doppler shifted. In the fully
gapped superconductor in Eq.(\ref{sWave}) the spectrum becomes 
$E({\bf p})=\pm\sqrt{(p^2/ 2m -\mu)^2+|\Delta|^2}+ {\bf p}\cdot {\bf v}_s$.
When the flow velocity exceeds the Landau critical velocity, $v_L\approx
\Delta/p_F$ if $\Delta\ll \mu$, the positive branch crosses zero energy
level. Typically this leads to instability, but in some cases, for
example, in superfluid $^3$He-B, the superfluidity is not destroyed.
In this case the Landau critical velocity marks the quantum phase
transition at which  two Fermi surfaces with $E({\bf p})=0$ emerge in the
superfluid state ({\bf b}). Liquid remains superfluid, but the density  of
the fermionic states is nonzero due to Fermi surfaces. Due to that the
normal component of the liquid becomes nonzero even at $T=0$, as a result
the density of the superfluid component
$\rho_s$ (the prefactor in the superfluid current ${\bf j}_s=\rho_s {\bf
v}_s$) is reduced compared with its value
$\rho$ below the threshold.
  See also Sec. 26.1 in Ref. \cite{VolovikBook}. }
\label{CritVelFig}
\end{figure}

 Similar Lifshitz transitions from the fully gapped
state to the state with the Fermi surface may occur in superfluids and
superconductors. This happens, for example,  when the superfluid velocity
crosses the Landau critical velocity [Fig. \ref{CritVelFig}].  The symmetry of the order parameter does not change
across such a quantum phase transition. On the other examples of the Fermi
surface in  superfluid/superconducting states in condensed matter and
quark matter see
\cite{Gubankova}).    In the non-superconduting states, the transition 
from the gapless to gapped state is the metal-insulator transition. The
Mott transition also belongs to this class.

\subsubsection{Reconnection of vortex lines in ${\bf p}$-space}

\begin{figure}[t]
\centerline{\includegraphics[width=\linewidth]{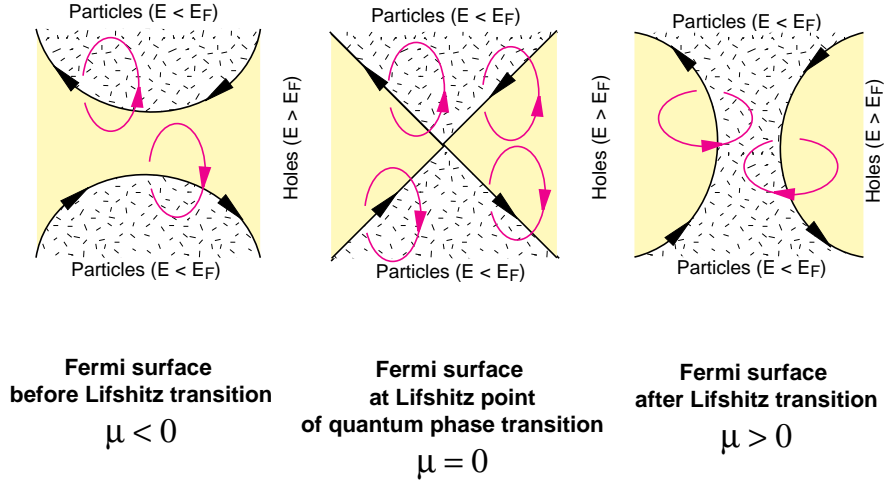}}
%\centerline{\epsfxsize=0.70\textwidth\epsfbox{Reconnection.eps}}
\medskip
\caption{  Lifshitz transition with change of the Fermi surface  topology
as reconnection of vortex lines in momentum space. The
direction of the "circulation" around  the vortex lines ({\it grey arrows})
and "vorticity" along  the vortex lines  ({\it black arrows}) are shown}
\label{ReconnectionFig}
\end{figure}

The Lifshitz transitions involving the  vortex lines  in ${\bf p}$-space
 may occur  between the gapless states. They are accompanied by the change of the topology of the
Fermi surface itself. The simplest example of such a phase transition discussed
in terms of the vortex lines  is provided by the
reconnection of the vortex lines. In Fig.
\ref{ReconnectionFig}  the two-dimensional system is  considered with the
saddle point spectrum  $E({\bf p})=p_x^2-p_y^2 -\mu$. The reconnection
quantum transition occurs at $\mu=0$. The three-dimensional systems, in
which the Fermi surface is a 2D vortex sheet in the 4D space
$(\omega,p_x,p_y,p_z)$, may experience the more complicated
topological transitions.

\subsection{Metal-superconductor transition} 
\label{Superconductivity}

The transition to superconducting state,  even if it occurs at $T=0$,
does not belong to the class of the quantum phase transitions which we
discuss in this review, because it is the consequence of the
spontaneously broken symmetry and does not occur perturbatively. Let us
discuss this transition from the point of view of the momentum-space
topology.

\subsubsection{Topology of Gor'kov function across  the superconducting
transition} 
\label{SuperconductivityTopChange}

Let us first note that the breaking of $U(1)$ symmetry is  not the
sufficient condition for superfluidity or  superconductivity.  For
example, the  $U(1)$ symmetry of the atoms A which is the result of
conservation of the number $N_A$ of A atoms, may be  violated simply due
to possibility of decay of atom A to atom B. But this does not lead to
superfluidity, and the Fermi surface  does not disappear. For these two
species of atoms  the Hamiltonian is $2\times 2$ matrix, such as 
\begin{equation}   
H=  
\left(\matrix{ p^2/ 2m_A  -\mu
&\Delta\cr
   \Delta^*&p^2/ 2m_B-\mu \cr }\right)~,
\label{mixing}
\end{equation}
where $\Delta$ is the matrix element which mixes the atoms A and B.  This
mixing violates the separate $U(1)$ symmetry for each of the two gases,
but the gap does not appear. Zeroes of the energy spectrum found from the
nullification of the determinant of the matrix, $(p^2/ 2m_A  -\mu)( p^2/
2m_B -\mu)- |\Delta|^2=0$, form two Fermi surfaces if $\Delta=0$, and
these Fermi surfaces survive if $\Delta\neq 0$ but is sufficiently small.
This is the consequence of topological stability of ${\bf p}$-space
vortices. Each Fermi surface has topological charge $N_1=1$, and their
sum $N_1=2$ is robust to small perturbations.

The non-perturbative phenomenon of superfluidity in the fermionic  gas
occurs due to Cooper pairing of atoms (electrons), i.e. due to mixing
between the particle and hole states.  Such mixing requires introduction
of the extended matrix Green's function even for a single fermions
species. This is the Gor'kov Green's function which is the matrix in the
particle-hole space of the same fermions, i.e. we have effective doubling
of the relevant fermionic degrees of freedom for the description of
superconductivity.  In case of $s$-wave pairing the Gor'kov Green's
function has the following form:
\begin{equation}   
G^{-1}=  
\left(\matrix{i\omega- p^2/ 2m+\mu
&\Delta\cr
   \Delta^*&i\omega +p^2/ 2m-\mu \cr }\right)~,
\label{Gorkov}
\end{equation}
Now the energy spectrum 
\begin{equation}   
E^2=(p^2/ 2m -\mu)^2+|\Delta|^2
\label{sWave}
\end{equation}
 has a gap, i.e. the Fermi surface disappears. How does this happen?
At $\Delta=0$ the matrix  Green's function describes two species of 
fermions: particles and holes. The topological charges of the
corresponding Fermi surfaces are $N_1=1$ for particles and $N_1=-1$ for
holes, with total topological charge $N_1=0$. The trivial total
topological charge of the Fermi surfaces allows for their annihilation,
which just occurs when the mixing matrix element $\Delta\neq 0$ and the
energy spectrum becomes fully gapped.  Thus the topology of the matrix
Gor'kov Green's function $G$ does not change across the superconducting
transition. 

 \subsubsection{Topology of diagonal Green's function across the 
superconducting transition} 
\label{DiagonalTopChange} 

Let us consider what happens with  the conventional Green's  function
across the transition. This is the $G_{11}$ element of the matrix
(\ref{Gorkov}):
\begin{equation}   
G_{11}=  -
\frac{i\omega + p^2/ 2m-\mu}{\omega^2+(p^2/ 2m -\mu)^2+|\Delta|^2}~.
\label{Gorkov11}
\end{equation}
One can see that it has the same topology in momentum space as the 
Green's function of normal metal  in Eq.(\ref{Propagator2}):
\begin{equation}   
G_{11}(\Delta=0)=  
\frac{1}{i\omega -p^2/ 2m+\mu}=-  
\frac{i\omega + p^2/ 2m-\mu}{\omega^2+(p^2/ 2m -\mu)^2}~.
\label{Gorkov11normal}
\end{equation}
Though instead of the pole in Eq.(\ref{Gorkov11normal}) for 
superconducting state one has zero in Eq.(\ref{Gorkov11}) for normal
state, their  topological charges  in Eq.(\ref{InvariantForFS}) are the
same: both have the same vortex singularity  with $N_1=1$. Thus the
topology of the conventional Green's function $G_{11}$ also does not
change across the superconducting transition. 

So the topology of each of the functions $ G$ and $G_{11}$ does  not
change across the transition. This illustrates again the robustness of
the topological charge. But what occurs at the transition?
 The Green's function $G_{11}$ gives the proper description of the 
normal state, but it does not provide the complete description of the
superconducting state, That is why its zeroes, though have non-trivial
topological charge, bear no information on the  spectrum of excitations.
On the other hand the matrix Green's function $G$ provides the complete
description of the superconducting states, but is meaningless on the
normal state side of the transition. Thus the spectrum on two sides of
the transition is determined by two different functions with different
topological properties. This illustrates the non-perturbative  nature of
the superconducting transition, which crucially changes the ${\bf
p}$-space topology leading to the destruction of the Fermi surface
without conservation of the topological charge across the transition.

\subsubsection{Momentum space topology in pseudo-gap state} 
\label{PseudoGap} 

Pseudo-gap is the effect of the suppression of the density  of states
(DOS) at low energy \cite{Pseudogap}.  Let us consider a simple model in
which the pseudo-gap behavior of the normal Fermi liquid results from the
superfluid/superconducting fluctuations, i.e.  in this model the
pseudo-gap state is the normal (non-superconducting) state with the
virtual superconducting order parameter $\Delta$  fluctuating about its
equilibrium zero value  (see review \cite{Sadovskii} and Ref.
\cite{Sadovskii2}). For simplicity we discuss the extreme case of such
state where $\Delta$  fluctuates being homogeneous in space.  The average
value of the off-diagonal element of the Gor'kov functions is zero in this
state,
$\left<G_{12}\right>=0$, and thus the $U(1)$ symmetry remains unbroken.
The Green's function of this pseudo-gap state is obtained by averaging of
the function $G_{11}$  over the distribution of the uniform complex order
parameter $\Delta$:
\begin{equation}    
G=\left<G_{11}\right>= \int d\Delta d\Delta^*P(|\Delta|) \frac {-i\omega
-\epsilon}{\omega^2 +\epsilon^2 + |\Delta|^2}~~.
\label{G11Pseudo}
\end{equation}
Here $\epsilon ({\bf p})=p^2/2m -\mu$ and $P(|\Delta|)$ is  the
probability of the gap $|\Delta|$. If $P(0)\neq 0$, then in the
low-energy limit $\omega^2 +\epsilon^2\ll
\Delta_0^2$, where $\Delta_0$ is the amplitude of fluctuations, one obtains
\begin{equation}    
G=  \frac{Z} {i\omega
-\epsilon}~~,~~ Z\propto \frac{\omega^2 +\epsilon^2}{\Delta_0^2}  \ln
\frac{\Delta_0^2}{\omega^2 +\epsilon^2}~.
\label{Z}
\end{equation}
The Green's function has the same topological property as  conventional
Green's function of metal with Fermi surface at $\epsilon({\bf p})=0$,
but the suppression of residue $Z$ is so strong, that the pole in the
Green's function  is transformed to the zero of the Green's function.
Because of the topological stability, the singularity of the Green's
function at the Fermi surface is not destroyed: the zero is also the
singularity and it has the same topological invariant  in
Eq.(\ref{InvariantForFS}) as pole. So this model of the Fermi liquid
represents a kind of Luttinger or marginal Fermi liquid with a very
strong renormalization of the singularity at the Fermi surface.
Transformation of poles to zeroes at the Mott transition has been
 discussed in Refs. \cite{Dzyaloshinskii,Phillips}. 

This demonstrates that the topology of the Fermi surface is the  robust
property, which does not resolve between different fine structures of the
Fermi liquids with different DOS. 

Using the continuation of Eq.(\ref{Z}) to the real frequency axis  
$\omega$, one obtains the density of states in this extreme model of the
pseudo-gap:  
\begin{equation}    
\nu(\omega)=N_0\int d\epsilon ~{\rm Im} G=\pi N_0\int_0^\omega
d\epsilon~\frac{\omega +\epsilon}{\Delta_0^2}= \frac {3\pi}{2}  N_0\frac
{\omega^2}{\Delta_0^2} ~~,
\label{DOS}
\end{equation}
where $N_0$ is the DOS of the conventional Fermi liquid,  i.e. without
the pseudo-gap effect. Though this state is non-superfluid and is
characterized by the Fermi surface,  the DOS at $\omega\ll \Delta_0$ is
highly suppressed compared to $N_0$, i.e. the pseudo-gap effect is highly
pronounced. This DOS has the same dependence on
$\omega$ as that in such superconductors or superfluids in which  the gap
has point nodes discussed  in the next Section \ref{FermiPoints}. When
the spatial and time variation of the gap fluctuations are taken into
account, the pseudo-gap effect would not be so strong.

\subsubsection{Momentum space topology in marginal Fermi liquid} 
\label{MarginalGL}

Actually in all the metals the Landau Fermi-liquid picture is violated
due to interaction of electrons with electromagnetic field
\cite{Holstein,Reizer}. The Green's function at very low frequency is
non-analytic and cannot be expressed  in terms of the pole: 
\begin{equation}    
G^{-1} = i\omega\left(1+\gamma \ln \frac {M^2}{\omega^2}\right)
-\epsilon~,~\gamma=  \frac{   e^2 v_F}{24 \pi^2
c^2}.
\label{Reizer}
\end{equation}
The same logarithmical divergence takes place for the fermionic Green's function
in the high-density quantum chromodynamics
due to interaction  with gluons
\begin{equation}    
\gamma=C_F\frac{g^2v_F}{24\pi^2} ~,~C_F=\frac{N_c^2-1}{2N_c}~,
\label{gamma}
\end{equation}
where $N_c(=3)$ is the number of colors  \cite{NonFLQCD}. 
Nevertheless the singularity  of the Green's function at the Fermi
surface (i.e. at $\omega=\epsilon=0$) is
described  by the same topological invariant $N_1$ in
Eq.(\ref{InvariantForFS}) as in the case of the conventional Fermi surface for the
conventional Green's function with poles.
The same topology remains for marginal Fermi liquids emerging
in 1D and 2D systems.
For example, the  Green's function of the 2D fermions interacting with the gauge-like bosons
 (see Ref. \cite{Khveshchenko} and references therein) 
\begin{equation}    
G^{-1} = i\omega\left( \frac {M^2}{\omega^2}\right)^{1/6}
-\epsilon~,
\label{2DReizer}
\end{equation}
is also described  by the Fermi surface  topological invariant $N_1$ in
Eq.(\ref{InvariantForFS}).

\section{Fermi points} 
\label{FermiPoints}
\subsection{Fermi point as topological object}

\subsubsection{Chiral Fermi points}

The crucial non-perturbative reconstruction of the spectrum occurs  at
the superfluid transition to
$^3$He-A, where the point nodes emerge instead of the Fermi surface. 
Since we are only interested in effects determined by the topology and
the symmetry of the fermionic Hamiltonian $H({\bf p})$ or Green's function
$G({\bf p},i\omega)$, we do not  require a special
form of the Green's function and can choose the simplest one
with the required topology and symmetry. First, consider the
Bogoliubov--Nambu Hamiltonian which qualitatively describes fermionic
quasiparticles in the axial state of $p$--wave pairing.
This Hamiltonian can be applied  to superfluid $^3$He-A 
\cite{VollhardtWoelfle} and 
also to the $p$-wave 
BCS state of ultracold Fermi gas:
\begin{eqnarray}   
H=  
\left(\matrix{ p^2/ 2m  -\mu
&c_\perp\,{\bf p}\cdot (\hat{\bf e}_1+ i\, \hat{\bf e}_2) \cr
    c_\perp\,{\bf p}\cdot (\hat{\bf e}_1- i\, \hat{\bf e}_2)
&-p^2/ 2m +\mu \cr }\right)\nonumber\\
= \tau_3(p^2/ 2m  -\mu)+ c_\perp\,{\bf p}\cdot (\tau_1\hat{\bf
e}_1- \tau_2 \hat{\bf e}_2),
\label{BogoliubovNambuH}
\end{eqnarray}
where $\tau_1$, $\tau_2$ and $\tau_3$ are $2\times 2$ Pauli matrices in
Bogoliubov--Nambu particle-hole space, and we neglect the spin structure
which is irrelevant for consideration. 
The orthonormal triad $(\hat{\bf e}_1,\, \hat{\bf e}_2,\,
\hat{\bf l}\equiv \hat{\bf e}_1\times \hat{\bf e}_2)$ characterizes
the order parameter in the axial state of triplet
superfluid.
 The unit vector $\hat{\bf l}$ corresponds to the
direction of the orbital momentum of the Cooper pair (or the
diatomic molecule in case of BEC); and
$c_\perp$ is the speed of the quasiparticles if they propagate
in the plane perpendicular to $\hat{\bf l}$. 

\begin{figure}[t]
\centerline{\includegraphics[width=0.90\linewidth]{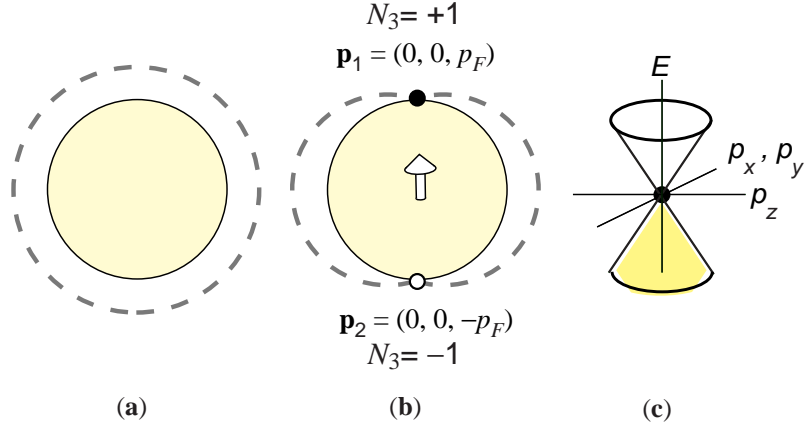}}
%\centerline{\epsfxsize=0.70\textwidth\epsfbox{FermiPoints.eps}}
\medskip
\caption{ Angular dependence of the superfluid/superconducting gap ({\it
dashed lines}) at the former Fermi surface ({\it solid lines}). The gap
$\Delta$ is  ({\bf a}) isotropic  in the $s$-wave state and  ({\bf b}) is
$\Delta(\theta)=p_Fc_\perp\sin\theta$ for the $p$-wave state in
Eq.(\ref{BogoliubovNambuE}), where $\theta$ is the polar angle, and arrow
shows the direction of the unit vector $\hat {\bf l}$. The spectrum of
quasiparticles has two nodes at the former Fermi surface: at 
$\theta=0$, i.e. at ${\bf p}_1=p_F\hat {\bf l}$ ({\it filled circle}) and 
at 
$\theta=\pi$, i.e. at ${\bf p}_2=-p_F\hat {\bf l}$ ({\it open circle}).
Their winding numbers of the Fermi points given by  
Eq.(\ref{TopInvariant}) are correspondingly
$N_3=+1$ and $N_3=-1$. ({\bf c}) According to Eq.(\ref{RelativisticE}),
close to the Fermi points the quasiparticle spectrum becomes
``relativistic''. For the local observer,  
who measures the spectrum using clocks and rods made of the
low-energy fermions, the Hamiltonian for fermions in the vicinity of the
point with $N_3=+1$ is equivalent to the Weyl Hamiltonian for the right
handed massless fermions:
$H= c{\mbox{\boldmath$\sigma$}}\cdot{\bf p}$; and the spectrum has the
conical form $E=\pm cp$
 }
\label{FermiPointsFig}
\end{figure}

The energy spectrum of these  Bogoliubov--Nambu fermions is
\begin{equation}
E^2 ({\bf p}) = \left({p^2\over 2m}-\mu\right)^2+\,
          c_\perp^2\,\left({\bf p}\times \hat{\bf l}\,\right)^2 .
\label{BogoliubovNambuE}
\end{equation}
In the BCS regime occuring for positive chemical potential $\mu>0$, there
are two Fermi points in 3D momentum space with $E({\bf p})=0$. For
the energy spectrum (\ref{BogoliubovNambuE}), the Fermi points are 
${\bf p}_1= p_F \, \hat{\bf l}$ 
and ${\bf p}_2=-p_F \, \hat{\bf l}$, with Fermi momentum
$p_F=\sqrt{2 m \mu}$ [Fig. \ref{FermiPointsFig}({\bf b})].

For a general system, be it relativistic or nonrelativistic, the
topological stability of the Fermi point (the node of the co-dimension 3)
is guaranteed by the nontrivial homotopy group $\pi_2(GL(n,{\bf C}))=Z$
which describes the mapping of a sphere $S^2$ embracing the point node to
the space  of non-degenerate complex matrices \cite{Horava}. This is
the group of integers. The integer valued  topological invariant (winding
number) can be written in terms of the fermionic propagator
$G(i\omega,{\bf p})$  as a surface integral in
the 4D frequency-momentum space $p_\mu=(\omega,{\bf p})$:
\cite{VolovikBook}
\begin{equation}    
N_3 \equiv {1\over{24\pi^2}}\,\epsilon_{\mu\nu\rho\sigma}~
{\rm tr} \,\oint_{\Sigma_a} dS^{\sigma}  
G\frac{\partial}{\partial p_\mu} G^{-1}\;
G\frac{\partial}{\partial p_\nu} G^{-1}\;
G\frac{\partial}{\partial p_\rho}G^{-1}.
\label{TopInvariant}
\end{equation}
Here $\Sigma_a$ is a three-dimensional surface around the
isolated Fermi point $p_{\mu a}=(0,{\bf p}_a)$ and `tr' stands for 
the trace over the relevant spin and/or band indices.
For the case considered in Eq.(\ref{BogoliubovNambuH}),  the Green's 
function  is $G^{-1}(i\omega,{\bf p})=i\omega - H({\bf p})$; the trace 
is over the Bogoliubov-Nambu spin; and the
two Fermi points  ${\bf p}_1$ and  ${\bf p}_2$
have nonzero topological charges $N_3=+1$ and
$N_3=-1$ [Fig. \ref{QuantPTransitionFig} ({\it right})]. 

We call such Fermi points the chiral Fermi
points, because in the vicinity of these point the fermions behave as 
right-handed or left handed particles (see below). These nodes of
co-dimension 3 are the diabolical points -- the exceptional degeneracy 
points  of the complex-valued Hamiltonian which depends on the external
parameters  (see
 Ref. \cite{NeumannWigner,StoneA,Arnold,Berry}). At these points two
different branches of the spectrum touch each other.  Topology of these
points has been discussed in Ref. \cite{Novikov}. In our case the
relevant parameters of the Hamiltonian are the components of momentum
${\bf p}$, and we discuss the contact point of branches with positive and
negative energies \cite{VolovikDiabolic}. Topology of the chiral Fermi
points in relation to the spectrum of elementary particles has been
discussed in Ref. \cite{NielsenNinomiya}.

\subsubsection{Emergent relativity and chiral fermions}

 Close to any of the Fermi points the energy spectrum
of fermionic quasiparticles acquires the relativistic form (this follows
from the so-called Atiyah-Bott-Shapiro construction \cite{Horava}).  In
particular, the Hamiltonian in Eq.(\ref{BogoliubovNambuH}) and 
spectrum in Eq.(\ref{BogoliubovNambuE}) become 
\cite{VolovikBook}:
\begin{equation}
H\rightarrow e_k^i \sigma^k(p_i- eA_i)~~,~~E^2 ({\bf p})
\rightarrow g^{ik}(p_i-eA_i) (p_k-eA_k)~ .
\label{RelativisticE}
\end{equation}
Here the analog of the dynamic gauge field is ${\bf A}=p_F\hat{\bf l}$; 
the ``electric charge''  is either $e=+1$ or $e=-1$ depending on the
Fermi  point; the matrix $e_i^k $ is the analog of the dreibein with 
$g^{ik}=e^i_j e^k_j={\rm
diag}(c_\perp^2,c_\perp^2,c_\parallel^2=p_F^2/m^2)$ playing the role of
the effective dynamic metric in which fermions move along the geodesic
lines. Fermions in Eq.(\ref{RelativisticE}) are chiral: they are
right-handed if the determinant of the matrix $e^i_j$ is positive, which
occurs at $N_3=+1$; the fermions are left-handed if the determinant  of
the matrix
$e^i_j$ is negative, which occurs at $N_3=-1$. For the local observer,  
who measures the spectrum using clocks and rods made of the
low-energy fermions, the Hamiltonian in Eq.(\ref{RelativisticE}) is
simplified: $H=\pm c{\mbox{\boldmath$\sigma$}}\cdot{\bf p}$ [Fig.
\ref{FermiPointsFig}({\bf c}).  Thus the chirality is the property of the
behavior in the low energy corner and it is determined by the topological
invariant
$N_3$.

\subsubsection{Majorana Fermi point}

The Hamiltonians which give rise to the chiral Fermi points with  non-zero
$N_3$ are essentially complex  matrices. That is why one may expect that
in systems described by real-valued  Hamiltonian matrices there are no
topologically stable points of co-dimension 3. However, the general
analysis in terms of $K$-theory \cite{Horava}
 demonstrates that such points exist and are described by the group $Z_2$.
 Let us denote this $Z_2$ charge as $N_{3{\rm M}}$ to distinguish
it from  the $Z$ charge $N_3$ of chiral fermions.  
The summation law for the charge $N_{3{\rm M}}$ is $1+1=0$, i.e. two
such points annihilate each other.  Example of topologically stable
massless real fermions   is provided by the Majorana fermions
\cite{Horava}.
The summation law
$1+1=0$ also means that $1=-1$, i.e. the particle is its own
antiparticle. This property of the Majorana fermions follows from the
topology in momentum space and does not require the relativistic
invariance. 

\subsubsection{Summation law for Majorana fermions  and marginal Fermi
point}

The summation law $1-1=0$ for chiral fermions and $1+1=0$ for  Majorana
fermions is illustrated using the following $4\times 4$ Hamiltonian
matrix:
\begin{equation}
H=c\tau_1 p_x +c\tau_2 \sigma_2 p_y + c\tau_3p_z~.
\label{Majorana}
\end{equation}
This Hamiltonian describes either two chiral fermions or  two Majorana
fermions. The first description is obtained if one chooses the spin
quantization axis along $\sigma_2$. Then for the direction of spin
$\sigma_2=+1$ this Hamiltonian describes the right-handed fermion with
spectrum $E(p)=cp$ whose Fermi point at ${\bf p}=0$ has topological
charge $N_3=+1$. For $\sigma_2=-1$ one has the left-handed chiral fermion
whose Fermi point is also at ${\bf p}=0$, but it has  the opposite
topological charge $N_3=-1$. Thus the total topological charge of the
Fermi point at ${\bf p}=0$ is $N_3=1-1=0$. 

In the other description, one takes into account that the  matrix
(\ref{Majorana}) is real and thus can describe the real (Majorana)
fermions. In our case the original fermions are complex, and thus we
have  two real fermions with the spectrum $E(p)=cp$ representing the real
and imaginary parts of the complex fermion. Each of the two Majorana
fermions has the Fermi (Majorana) point at ${\bf p}=0$ where the energy
of fermions is zero. Since the Hamiltonian (\ref{Majorana}) is the same
for both real fermions, the two Majorana points have the same topological
charge.

Let us illustrate the difference in the summation law for  charges $N_3$
and $N_{3{\rm M}}$ by introducing  the perturbation $M\sigma_1\tau_2$ to
the Hamiltonian (\ref{Majorana}):
\begin{equation}
H=c\tau_1 p_x +c\tau_2 \sigma_2 p_y + c\tau_3p_z +M\sigma_1\tau_2~.
\label{DiracM}
\end{equation}
Due to this perturbation the spectrum  of fermions is fully gapped: 
$E^2(p)=c^2p^2+M^2$. In the description in terms of the chiral fermions,
the perturbation mixes left and right fermions. This leads to  formation
of the Dirac mass $M$. The annihilation of  Fermi points with opposite
charges  illustrates the summation law $1-1=0$ for the topological charge
$N_3$.

Let us now consider the same process using the description in  terms of
real fermions. The added term $M\sigma_1\tau_2$ is imaginary. It mixes
the real and imaginary components of the complex fermions, and thus it
mixes two Majorana fermions.  Since the two  Majorana  fermions have the
same topological charge,  $N_{3{\rm M}}=1$, the formation of the gap
means that the like charges of the Majorana  points annihilate each
other. This illustrates the summation law $1+1=0$ for the Majorana 
fermions. 

In both descriptions of  the Hamiltonian (\ref{Majorana}),  the total
topological charge of the Fermi or Majorana point at ${\bf p}=0$ is zero.
We call such topologically trivial point the marginal Fermi point. The
topology does not protect the marginal Fermi point, and the small
perturbation can lead to formation of the fully gapped vacuum, unless
there is a symmetry which prohibits this.

\subsection{Quantum phase transition in BCS--BEC crossover region} 

\subsubsection{Splitting of marginal Fermi point}

Let us consider some examples of  quantum phase transition
goverened by the momentum-space  topology of gap nodes, 
 between a  fully-gapped vacuum state and a vacuum state
with topologically-protected point nodes. In the
context of condensed-matter physics, such a quantum phase
transition may occur in a system of
ultracold fermionic atoms in the region of the BEC--BCS
crossover, provided Cooper pairing occurs in the non-$s$-wave
channel. For elementary particle physics, such transitions are
 related to CPT violation, neutrino
oscillations, and other phenomena \cite{SplittingPreprint}.

\begin{figure}[t]
\centerline{\includegraphics[width=0.90\linewidth]{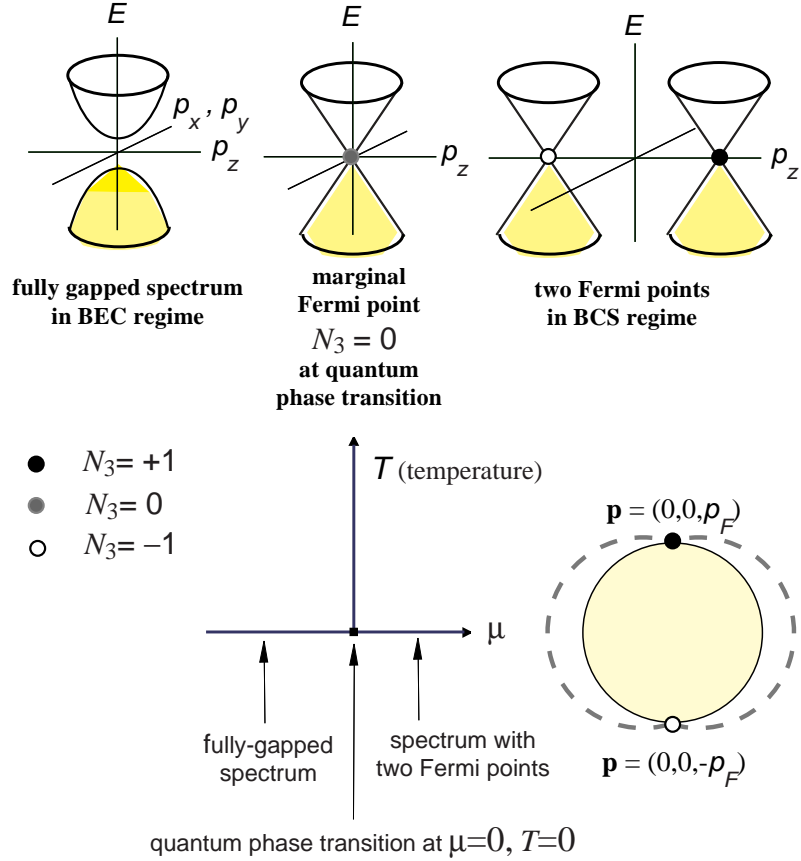}}
%\centerline{\epsfxsize=0.70\textwidth\epsfbox{BCSBEC.eps}}
\medskip
\caption{Quantum phase transition between two $p$-wave vacua with the same
symmetry but of different universality class.
In the bottom right corner you find the gap 
({\it dashed line}) in the $p$-wave state. It depends on the
direction in momentum space and becomes zero when ${\bf p}$ is
along the $\hat{\bf l}$-vector (along
$z$-axis). At $\mu>0$, two gap nodes give rise to two zeroes in the
spectrum -- Fermi points:  one with
winding number
$N_3=+1$  ({\it filled circle}) and another with
winding number $N_3=-1$  ({\it open circle}).   The transition occurs when
the chemical potential
$\mu$ in Eq.(\ref{BogoliubovNambuH}) crosses
zero value.
The Fermi points merge at
$\mu=0$ forming the  marginal (topologically trivial) gap node with
$N_3=0$  ({\it grey circle}) and  annihilate each other. At $\mu<0$ the
Green's function has no singularities and the  quantum vacuum is fully
gapped  
 }
\label{QuantPTransitionFig}
\end{figure}

Let us start with the topological quantum phase transition  involving 
topologically stable Fermi points \cite{Book1,QPT}.  Let us consider what
happens with the Fermi points in   Eq. (\ref{BogoliubovNambuE}), when one
varies the chemical potential $\mu$. For $\mu>0$, there are two Fermi
points, and the density of  fermionic states in the vicinity of  Fermi
points is $\nu(\omega)\propto \omega^2$. For
$\mu<0$,  Fermi points are absent and the spectrum is fully-gapped [Fig.
\ref{QuantPTransitionFig}]. In this topologically-stable fully-gapped
vacuum, the density of states is drastically different from that in the
topologically-stable gapless regime: $\nu( \omega)=0$ for
$ \omega<|\mu|$. This demonstrates that
the quantum phase transition considered is of purely topological origin.
The transition occurs at $\mu=0$, when two Fermi points with $N_3=+1$
and  $N_3=-1$ merge and form one topologically-trivial Fermi point with
$N_3=0$, which disappears at $\mu<0$. 

The intermediate state at $\mu=0$ is
marginal: the momentum-space topology is trivial ($N_3=0$) and  cannot
protect the vacuum against decay into one of the two topologically-stable
vacua unless there is a special symmetry which stabilizes the marginal
node. As we shall see in the Sec. \ref{StandardModel}, the latter takes
place in the Standard Model with marginal Fermi point. 

\subsubsection{Transition involving multiple nodes}

The Standard Model contains 16 chiral fermions in each generation. The
multiple Fermi point may occur in condensed matter too. For
systems of cold atoms, an example is provided by another  spin-triplet
$p$--wave state, the so-called 
$\alpha$--phase.  The 
Bogoliubov-Nambu Hamiltonian which
qualitatively describes fermionic quasiparticles in the 
$\alpha$--state is given by \cite{VolovikGorkov1985,VollhardtWoelfle}:
\begin{equation}
H= \left(\matrix{ p^2/ 2m  -\mu
&   \left( {\bf \Sigma} \cdot {\bf p}\right)        \,c_\perp/\sqrt{3} \cr
    \left( {\bf \Sigma} \cdot {\bf p}\right)^\dagger\,c_\perp/\sqrt{3}  
&-p^2/ 2m +\mu \cr } \right),
\label{BogoliubovNambuAlphaH}
\end{equation}
with  ${\bf \Sigma} \cdot {\bf p}\equiv 
\sigma_x p_x + \exp(2\pi i/3)\,\sigma_y p_y  +
\exp(-2\pi i/3)\,\sigma_z p_z\,$.  

\begin{figure}[t]
\centerline{\includegraphics[width=0.80\linewidth]{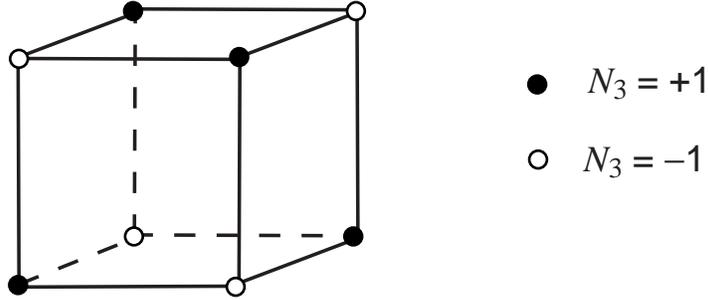}}
%\centerline{\epsfxsize=0.70\textwidth\epsfbox{Cube.eps}}
\medskip
\caption{Fermi points in the $\alpha$-phase of triplet
superfluid/superconductor in the BCS regime}
\label{QuantPTransitionFig3}
\end{figure}

On the BEC side ($\mu<0$), fermions are again fully-gapped, while on
the BCS side ($\mu>0$), there are 8 topologically protected
Fermi points with charges $N_3=\pm 1$,  situated at the vertices of a
cube in momentum space \cite{VolovikGorkov1985} [Fig.
\ref{QuantPTransitionFig3}]. The fermionic excitations in the vicinity of
these points are left- and right-handed Weyl fermions. At the transition
point at $\mu=0$ these 8 Fermi points merge forming the marginal Fermi
point at ${\bf p}=0$.

\subsection{Quantum phase transitions in Standard Model}
\label{StandardModel}

\subsubsection{Marginal Fermi point in Standard Model} 

\begin{figure}[t]
\centerline{\includegraphics[width=0.80\linewidth]{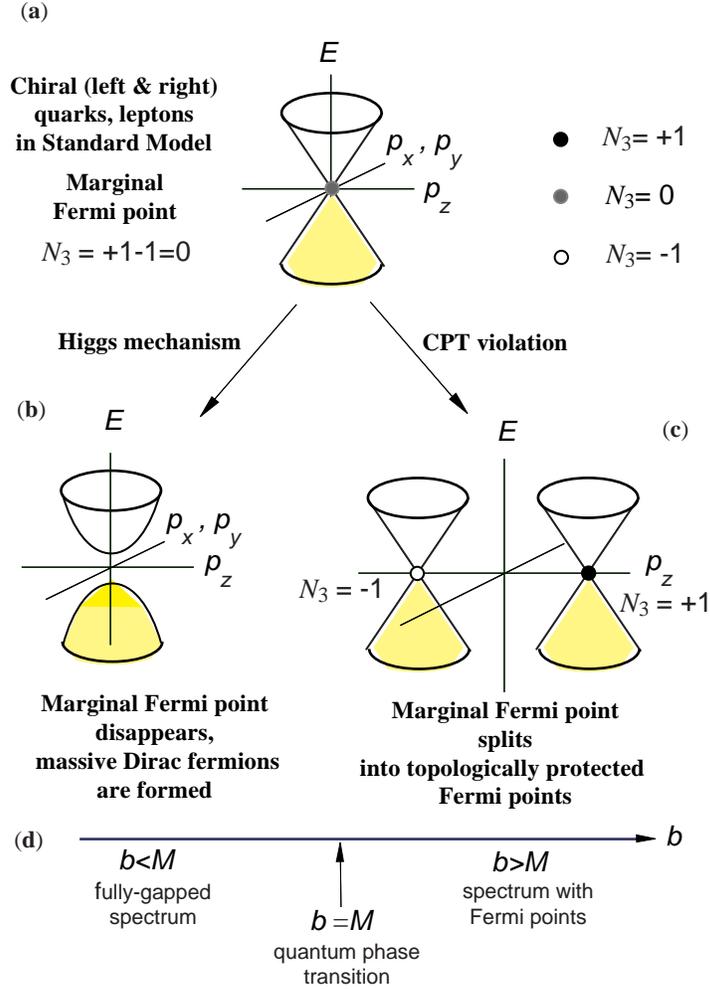}}
%\centerline{\epsfxsize=0.70\textwidth\epsfbox{TwoScenariaStandModel.eps}}
\medskip
\caption{Two scenarios of annihilation of marginal Fermi point ({\bf a})
in Standard Model of strong and electroweak interactions. Higgs mechanism
leads to Dirac mass and thus to the fully gapped vacuum ({\bf b}), while
CPT  violation leads to splitting of Fermi points ({\bf c}). In the bottom
edge you find the quantum phase transition in the model in
Eq.(\ref{ModifiedHDirac}) when the CPT violating parameter  $b\equiv|{\bf
b}|$ crosses  the Dirac mass $M$
}
\label{QuantPTransitionFig2}
\end{figure}

It is assumed that the Standard Model 
above the electroweak transition contains 16 chiral fermions  in each
generation: 8 right-handed fermions with $N_3=+1$ each and 8 left-handed
fermions with $N_3=-1$ each. If so, then the vacuum of the Standard
Model  above the electroweak transition is marginal:  there is a multiply
degenerate Fermi point at ${\bf p}=0$ with the total topological charge
$N_3=+8-8=0$ [Fig.~\ref{QuantPTransitionFig2}({\bf a})]. This vacuum is
therefore the intermediate state between two topologically-stable vacua:
the fully-gapped vacuum in Fig.~\ref{QuantPTransitionFig2}({\bf b}); and
  the vacuum with topologically-nontrivial Fermi points in
Fig.~\ref{QuantPTransitionFig2}({\bf c}). 

The absence of the topological
stability means that even  the small
 mixing between
the fermions leads to annihilation of the Fermi point.
 In the Standard Model, the proper mixing which leads to the fully gapped 
vacuum  is prohibited by symmetries, namely the continuous electroweak
$U(1)\times SU(2)$ symmetry (or the discrete symmetry discussed  in Sec.
12.3.2 of Ref.\cite{VolovikBook}) and the CPT symmetry. 
 (Marginal gapless fermions emerging in spin systems were discussed in
\cite{WenPRL}. These massless Dirac fermions protected  by symmetry
differ from the chiral fermions of the Standard Model. The latter cannot
be represented in terms of massless Dirac fermions, since there is no
symmetry between left and right fermions in Standard Model.)

Explicit violation or spontaneous breaking of electroweak or CPT symmetry
transforms the marginal vacuum of the Standard Model
into one of the two topologically-stable vacua.  If, 
for example, the electroweak symmetry is broken, the marginal Fermi point
disappears and the fermions become massive  
[Fig.~\ref{QuantPTransitionFig2}({\bf b})]. This is assumed to happen 
 below the
symmetry breaking electroweak transition caused by Higgs mechanism  where
quarks and charged leptons acquire the Dirac masses.  
 If, on the other hand, the CPT symmetry is violated, the
marginal Fermi point splits into topologically-stable Fermi
points which protect chiral fermions [Fig.~\ref{QuantPTransitionFig2}({\bf
c})].  One can speculate that in the  Standard Model the latter happens
with the electrically neutral leptons, the neutrinos
\cite{SplittingPreprint,KlinkhamerCPT}. 

\subsubsection{Quantum phase transition with splitting of Fermi points} 

Let us consider this scenario on a simple example of a marginal
Fermi point describing a \emph{single} pair of relativistic chiral
fermions, that is, one right-handed fermion and one left-handed
fermion. These are Weyl fermions with Hamiltonians $H_{\rm
right}={\mbox{\boldmath$\sigma$}}\cdot{\bf p}$ and $H_{\rm
left}=-{\mbox{\boldmath$\sigma$}}\cdot{\bf p}$,  where
${\mbox{\boldmath$\sigma$}}$ denotes the triplet of spin Pauli matrices.
Each of these Hamiltonians has a topologically-stable Fermi  point at
${\bf p}=0$.  The corresponding inverse Green's functions are given by
\begin{eqnarray} 
G^{-1}_{\rm right}(i\omega,{\bf p})&=&i\omega
-{\mbox{\boldmath$\sigma$}}\cdot{\bf p}\;,
\nonumber\\[2mm]
G^{-1}_{\rm left} (i\omega,{\bf p})&=&i\omega +{\mbox{\boldmath$\sigma$}}
\cdot{\bf
p}~.
\label{GreenFWeyl}
\end{eqnarray}
The positions of the Fermi points coincide, ${\bf p}_1={\bf
p}_2=0$, but their topological charges (\ref{TopInvariant}) are
different.  For this simple case, the topological charge equals
the chirality of the fermions, $N_3=C_a$ (i.e., $N_3=+1$ for the
right-handed fermion and $N_3=-1$ for the left-handed one).
The total topological charge of the Fermi point at ${\bf p}=0$ is
therefore zero.

The splitting of this marginal Fermi point can be described by
the Hamiltonians
$H_{\rm right}={\mbox{\boldmath$\sigma$}}\cdot({\bf
p}-{\bf p}_1)$ and
$H_{\rm left}=-{\mbox{\boldmath$\sigma$}}\cdot({\bf p}-{\bf
p}_2)$, with
${\bf p}_1=-{\bf p}_2 \equiv {\bf b}$ from momentum conservation.
The  real vector ${\bf b}$ is assumed to be odd under CPT,
which introduces CPT violation into the physics.
The $4\times 4$ matrix of the combined Green's function has the form
\begin{equation}  
G^{-1}(i\omega,{\bf p}) =
\left(\matrix{i\omega -
{\mbox{\boldmath$\sigma$}}\cdot({\bf p}-{\bf b})&0\cr 
0&i\omega+
{\mbox{\boldmath$\sigma$}}\cdot({\bf p}+{\bf b})\cr } \right).
\label{ModifiedGreenWeyl}
\end{equation}
    Equation ~(\ref{TopInvariant}) shows that ${\bf p}_1={\bf b}$ is
the Fermi point with topological charge $N_3=+1$ and
${\bf p}_2=-{\bf b}$
the Fermi point  with topological charge $N_3=-1$.

Let us now consider the more general situation with both the
electroweak and CPT symmetries broken. Due to breaking of the electroweak 
symmetry the Hamiltonian acquires the off-diagonal term (mass term) which
mixes left and right fermions
\begin{equation}
H =\left(\matrix{{\mbox{\boldmath$\sigma$}}\cdot({\bf p}-{\bf b})&M\cr
       M&-
{\mbox{\boldmath$\sigma$}}\cdot({\bf p}+{\bf b})\cr } \right)  ~.
\label{ModifiedHDirac}
\end{equation}
 The energy spectrum of Hamiltonian
(\ref{ModifiedHDirac}) is
\begin{equation}
E^2_\pm ({\bf p}) = M^2+|{\bf p}|^2+b^2\pm \, 2\,b\, \sqrt{M^2+\left({\bf
p}\cdot\hat{{\bf b}}\right)^2}~,
\label{ModifiedEnergyDirac}
\end{equation}
with $\hat{{\bf b}} \equiv {\bf b}/|{\bf b}|$ 
and $b\equiv |{\bf b}|$.

Allowing for a variable parameter $b$, one finds a quantum phase
transition at $b=M$ between the fully-gapped vacuum for $b<M$ and
the vacuum with two isolated Fermi points for $b>M$ 
[Fig.~\ref{QuantPTransitionFig2}({\bf d})]. These Fermi points are
situated at
\begin{eqnarray} 
{\bf p}_1 &=&+ \hat{{\bf b}}\;\sqrt{b^2-M^2}  \;,\nonumber\\[2mm]
{\bf p}_2 &=&- \hat{{\bf b}}\;\sqrt{b^2-M^2}   ~.
\label{FPDiracFermions}
\end{eqnarray}
        Equation ~(\ref{TopInvariant}),
now with a trace over the indices of the $4\times 4$ Dirac
matrices, shows that  the Fermi point at ${\bf p}_1$ has
topological charge $N_3=+1$ and thus the right-handed chiral fermions 
live in the vicinity of this point. Near the Fermi point at 
${\bf p}_2$ with the charge $N_3=-1$, the left-handed fermions live.   
The magnitude of the splitting of the two Fermi points is given by
$2\,\sqrt{b^2-M^2}\,$. At the quantum phase transition $b=M$,  the Fermi
points with opposite charge annihilate each other and form a marginal
Fermi point at
${\bf p} =0$. The momentum-space topology of this marginal Fermi  point
is trivial (the topological invariant $N_3=+1-1=0$).

\subsubsection{Fermi surface with global charge $N_3$ and quantum phase 
transition with transfer of $N_3$}
\label{FermiSurfaceGlobalCharge}  

\begin{figure}[t]
\centerline{\includegraphics[width=1.00\linewidth]{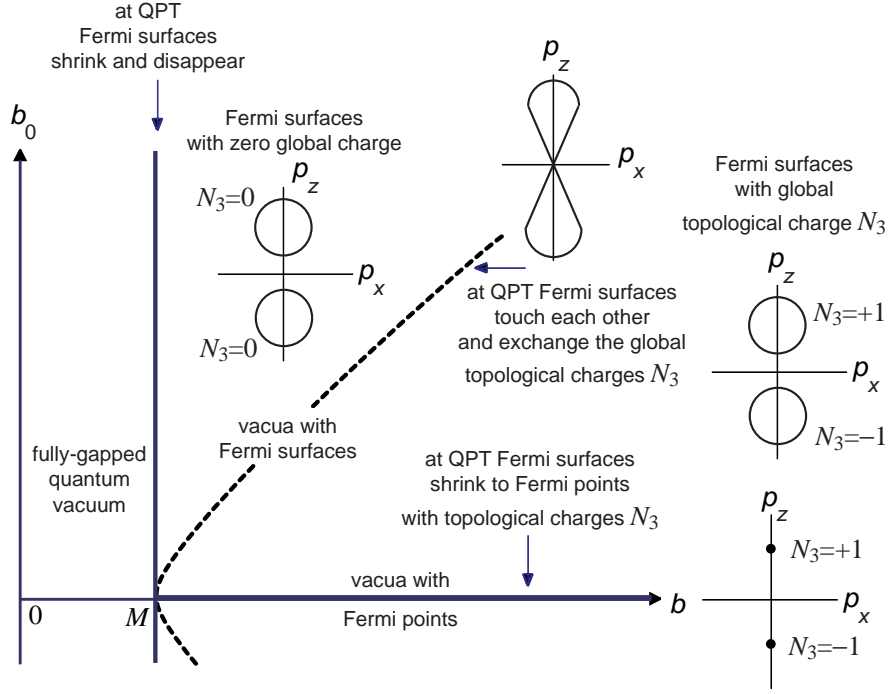}}
%\centerline{\epsfxsize=0.90\textwidth\epsfbox{FSTouch2.eps}}
\medskip
\caption{Topological quantum phase transitions in the model
(\ref{ExtModifiedHDirac}).  The vacua at $b_0\neq 0$ and $b>M$ have Fermi
surfaces. At $b^2>b_0^2+M^2$, these Fermi surfaces have nonzero global
topological  charges  $N_3=+1$ and  $N_3=-1$. At the quantum phase
transition occurring on the line $b_0=0$, $b>M$ ({\it thick horizontal 
line}) the Fermi surfaces shrink to the Fermi points with nonzero $N_3$.
At 
$M^2<b^2<b_0^2+M^2$ the global topology of the Fermi surfaces is
trivial, $N_3=0$.   At the quantum phase
transition occurring on the line   $b=M$ ({\it thick vertical 
line}),  the Fermi surfaces shrink to the points; and since their
global topology is trivial the zeroes disappear at
$b<M$ where the vacuum is fully gapped. The quantum phase transition
between the Fermi surfaces with and without topological charge $N_3$
occurs at $b^2=b_0^2+M^2$  ({\it dashed line}). At this transition, the
Fermi surfaces touch each other, and their topological charges annihilate
each other}
\label{FSTouchFig}
\end{figure}

Extension of the model (\ref{ModifiedHDirac}) by introducing the time 
like parameter $b_0$
\begin{equation}
H =\left(\matrix{{\mbox{\boldmath$\sigma$}}\cdot({\bf p}-{\bf b})-b_0&M\cr
       M&-
{\mbox{\boldmath$\sigma$}}\cdot({\bf p}+{\bf b})+b_0\cr } \right)  ~,
\label{ExtModifiedHDirac}
\end{equation}
demonstrates another type of quantum phase transitions
\cite{SplittingPreprint} shown in Fig. \ref{FSTouchFig}.

At $b_0\neq 0$, Fermi points which exist at $b_0=0$, $b>M$ transform to the
closed Fermi surfaces. These Fermi surfaces in addition to the  local
charge $N_1$ have the global topological invariant $N_3$ inherited from 
the original Fermi points. The global charge $N_3$  is defined by the same
Eq.~(\ref {TopInvariant}), but with a three-dimensional surface $\Sigma_a$
around  the whole Fermi  surface. On the line of the quantum phase
transition, 
$b^2-b_0^2=M^2$  ({\it dashed line}),  two Fermi surfaces contact each
other at the point ${\bf p}=0$. At that moment, the topological charge
$N_3$ is transferred between the Fermi surfaces through the point of the
contact. Above the transition line, the global charges of Fermi surfaces
are zero. At the quantum phase transition at $b=M$ ({\it thick vertical 
line}) these Fermi surfaces
shrink to the points; and since the $N_3$ topology of these points is
trivial they disappear at $b<M$ where the state is fully gapped.

\subsubsection{Standard Model with chiral Fermi point}

In the above consideration we assumed that the Fermi point in the
Standard Model above the electroweak energy scale is marginal, i.e. its
total topological charge is
$N_3=0$.  Since the topology does not protect such a point, everything
depends on symmetry, which is a more subtle issue. In principle, one may
expect that the vacuum is always fully gapped.  This is supported by the
Monte-Carlo simulations which suggest that in the Standard Model there is
no second-order phase transition at finite temperature, instead one has
either the first-order  electroweak transition or crossover depending on
the ratio of masses of the Higgs and gauge bosons 
\cite{Kajantie}. This would actually mean that the fermions are always
massive. 

Such scenario does not contradict to the
momentum-space topology, only if the total topological charge  
$N_3$ is zero. However, from the point of view of the momentum-space
topology there is another scheme of the description of the Standard
Model. Let us assume that the Standard Model follows from the  GUT with 
$SO(10)$ group.  In this scheme, the 16 Standard Model fermions  form at
high energy the 16-plet of the $SO(10)$ group. All the particles of this
multiplet are left-handed fermions. These are:  four left-handed $SU(2)$
doublets (neutrino-electron and 3 doublets of quarks) + eight  left
$SU(2)$ singlets of anti-particles (antineutrino, positron and 6
anti-quarks). The total topological charge of the   Fermi point at ${\bf
p}=0$ is $N_3=-16$, and thus such a vacuum is topologically stable and is
protected against the mass of fermions. This topological protection works
even if the $SU(2)\times U(1)$ symmetry is violated perturbatively, say,
due to the mixing of different species of the 16-plet. Mixing of left
leptonic doublet with left singlets (antineutrino and positron)  violates
$SU(2)\times U(1)$ symmetry, but this does not lead to annihilation of
Fermi points and mass formation since the topological charge $N_3$ is
conserved. 

We discussed the similar situation in the Sec.
\ref{Superconductivity} for the case of the Fermi surface, and found that
if the total topological charge of the Fermi surfaces is non-zero, the
gap cannot appear perturbatively. It can only arise due to the crucial
reconstruction of the fermionic spectrum with effective doubling of
fermions.  In the same manner, in the $SO(10)$ GUT model the mass
generation can only occur non-perturbatively. The mixing of the left and
right fermions requires the introduction of the right fermions, and thus
the effective doubling of the number of fermions.
The corresponding Gor'kov's Green's function in this case will be the
$(16\times 2)\times (16 \times 2)$ matrix. 
The nullification of the topological charge $N_3=-16$    occurs
exactly in the same manner, as in superconductors. In the extended
(Gor'kov) Green's function formalism appropriate below the transition, the
topological charge of the original Fermi point is annihilated by the
opposite charge
$N_3=+16$ of the Fermi point of `holes' (right-handed particles). 

This demonstrates that the mechanism of generation of mass of fermions
essentially depends on the momentum space topology.  If the Standard Model originates
from the $SO(10)$ group,  the vacuum
belongs to the universality class with the topologically non-trivial
chiral Fermi point (i.e. with $N_3\neq 0$), and the smooth crossover to 
the fully-gapped vacuum is impossible. On the other hand, if the Standard
Model originates from the left-right symmetric Pati-Salam group such as
$SU(2)_L\times SU(2)_R\times SU(4)$,  and its vacuum has  the
topologically trivial (marginal) Fermi point  with $N_3= 0$,  the smooth
crossover to the fully-gapped vacuum is possible.

\subsubsection{Chiral anomaly}

Since chiral Fermi points in condensed matter and in Standard Model are
described by  the same  momentum-space topology, one may expect common
properties. An example of such a common property would be the axial or
chiral anomaly. For quantum anomalies in (3+1)--dimensional systems with
Fermi points and their dimensional reduction to (2+1)--dimensional
systems,  see, e.g.,  Ref.~\cite{VolovikBook} and
references therein. In superconducting and superfluid fermionic systems
the chiral anomaly is instrumental for the dynamics of vortices. In
particular, one of the forces acting on continuous vortex-skyrmions in
superfluid $^3$He-A is the result the anomalous production of the
fermionic charge from the vacuum decsribed by the Adler-Bell-Jackiw
equation \cite{Adler}.

\section{Fermi lines}
\label{FermiLines}

In general the zeroes of co-dimension 2 (nodal lines in 3D momentum 
space or point nodes in 2D momentum space) do not have the topological
stability.  However, if the Hamiltonian is restricted by some symmetry,
the topological stability of these nodes is possible. The nodal lines do
not appear in spin-triplet superconductors, but they may exist in
spin-singlet superconductors
\cite{VolovikGorkov1985,Blount}. The analysis of topological stability 
of nodal lines in systems with real fermions was done by Horava
\cite{Horava}. 

\subsection{Nodes in high-$T_c$ superconductors}
\label{NodesSuperconductors}

An example of point nodes in 2D momentum space is provided  by the 
layered quasi-2D high-T$_c$ superconductor.   In the simplest form,
omitting the mass and the amplitude of the order parameter, the 2D
Bogoliubov-Nambu Hamiltonian is
 \begin{equation} 
H=\tau_3\left( \frac{p_x^2+ p_y^2}{2m}-\mu\right)+  a\tau_1(p_x^2-\lambda
p_y^2) ~. 
\label{dWaveHamiltonian}
\end{equation}
In case of tetragonal crystal symmetry one has either the pure $s$-wave
state with
$\lambda=-1$  ($p_x^2+ p_y^2$) or the pure $d$-wave state with
$\lambda=+1$  ($p_x^2- p_y^2$). But in case of orthorhombic crystal these
two states are not distinguishible by symmetry and thus the general 
order parameter is represented by the $s+d$ combination, i.e. in the
orthorhombic crystal one always has
$|\lambda|\neq 1$.  For example, experiments in high-T$_c$ cuprate 
YBa$_{2}$Cu$_3$O$_{7}$  suggest that
$\lambda\sim 0.7$ in this compound \cite{AnisotropyExperiment}.

At $\mu>0$ and $\lambda>0$, 
the energy spectrum contains 4 point nodes in 2D momentum space
(or four Fermi-lines in the 3D momentum space):
\begin{equation} 
p_x^a=\pm p_F\sqrt{\lambda\over 1+\lambda}~~,~~
p_y^a=\pm p_F\sqrt{1\over 1+\lambda}~~,~~ p_F^2=2\mu ~. 
\label{Nodes}
\end{equation}
%Here $a=++,+-,-+,--$. 

The problem is whether these nodes survive or not
if we extend Eq.(\ref{dWaveHamiltonian}) to the more general Hamiltonian
obeying the same symmetry.
The important property of this Hamiltonian is that, as distinct from
the Hamiltonian (\ref{BogoliubovNambuH}), it obeys the  time reversal
symmetry which prohibits  the imaginary $\tau_2$-term.
In the spin singlet states the Hamiltonian obeying the time reversal
symmetry must satisfy the equation
$H^*(-{\bf p})=H({\bf p})$.  The general
form of the $2\times 2$ Bogoliubov-Nambu spin-singlet Hamiltonian 
satisfying  this equation can be expressed in terms of the 2D vector
${\bf m}({\bf p})=(m_x({\bf p}),m_y({\bf p}))$:
 \begin{equation} 
H=\tau_3m_x({\bf p})+ \tau_1 m_y({\bf p}) ~. 
\label{2times2Hamiltonian}
\end{equation}
Using this vector one can construct 
the integer valued topological invariant --  the contour integral around
the point node in 2D momentum space or around the nodal line in 3D
momentum space: 
 \begin{equation} 
N_2=\frac {1}{2\pi} \oint dl ~\hat{\bf z} \cdot \left(\hat{\bf m} \times  
\frac{d\hat{\bf m}} {dl}\right)~, 
\label{Invariant}
\end{equation}
where $\hat{{\bf m}} \equiv {\bf m}/|{\bf m}|$. This is the winding 
number of the plane vector 
${\bf m}({\bf p})$ around 
a  vortex line  in 3D momentum space or around a point vortex in 2D
momentum space. The winding number is robust to any change of the
Hamiltonian respecting the time reversal symmetry, and this is the reason
why the node is stable.

All four nodes in the above example of Eq.(\ref{dWaveHamiltonian}) are
topologically stable, since nodes with equal signs (++ and $--$)  have
winding number
$N_2=+1$, while the other two nodes have winding
number
$N_2=-1$ [Fig. \ref{QuantPTransitionFig4}]. 
 
 \subsection{$Z_2$-lines}
\label{Z2lines}

Now let us consider the stability of these nodes using the  general 
topological analysis (the so-called $K$-theory, see \cite{Horava}). For
the general $n\times n$ real matrices the classification of the
topologically stable nodal lines  in 3D 
 momentum space (zeroes of co-dimension 2) is given by the homotopy 
group $\pi_1(GL(n,{\bf R}))$ \cite{Horava}. It  determines classes of
mapping of a contour $S^1$ around the nodal line (or around a point in
the 2D momentum space) to the space  of non-degenerate real matrices. The
topology of nodes depends on $n$. If $n=2$, the homotopy group for lines
of nodes is
$\pi_1(GL(2,{\bf R}))=Z$, it is the group of integers in
Eq.(\ref{Invariant})  obeying the conventional summation $1+1=2$.
However, for larger $n \geq 3$ the homotopy
group for lines of nodes is
$\pi_1(GL(n,{\bf R}))=Z_2$, which means that the summation law for the 
nodal lines is now $1+1 =0$, i.e. two nodes with like topological
charges annihilate each other.  These nodes of
co-dimension 2 are similar to the points of degeneracy of the 
energy spectrum of the real-valued Hamiltonian which depends on the
external parameters (see
 Ref. \cite{NeumannWigner,Arnold,Berry}).

The equation  (\ref{dWaveHamiltonian}) is  the $2\times 2$ Hamiltonian
for the complex fermionic field. But each complex field consists of two
real fermionic field. In terms of the real fermions, this Hamiltonian is 
the  $4\times 4$ matrix and thus all the nodes must be topologically
unstable. What keep them alive is the time reversal symmetry, which does
not allow to mix real and imaginary components of the complex field. As a
result, the two components are independent; they are described by the
same $2\times 2$ Hamiltonian  (\ref{dWaveHamiltonian}); they have zeroes
at the same points; and these  zeroes are described by the same
topological invariants. 

If we allow mixing between real and imaginary components  of the spinor
by introducing the imaginary perturbation to the Hamiltonian, such as 
$M\tau_2$, the summation law  $1+1$ leads to immediate annihilation of
the zeroes situated at the same points. As a result the spectrum becomes
fully gapped:
 \begin{equation} 
E^2({\bf p})= \left(\frac{p_x^2+ p_y^2}{2m}-\mu\right)^2+ a^2(p_x^2-
\lambda p_y^2)^2+ M^2 ~. 
\label{FullyGappedTimeIrreversal}
\end{equation}

Thus to destroy the nodes of co-dimension 2 occurring for $2\times$ 
real-valued Hamiltonian (\ref{dWaveHamiltonian}) describing  complex
fermions it is enough to violate the time reversal symmetry. 

How to destroy the nodes if the time reversal symmetry is  obeyed which
prohibits mixing? One possibility is to deform the order parameter in
such a way that the nodes with opposite $N_2$ merge and then annihilate
each other forming the fully gapped state.  In this case,  at the border
between the state with nodes and the fully gapped state the quantum phase
transition occurs (see Sec. \ref{TransitionNodalSuperconductor}). This
type of quantum phase transition which involves zeroes of co-dimension 2
was also discussed in Ref.\cite{WenZee}.

Another possibility is to increase the dimension of  the matrix from
$2\times 2$ to $4\times 4$. Let us consider this case.

\subsection{Gap induced by interaction between layers}

High-$T_c$ superconductors typically have several  superconducting
cuprate layers per period of the lattice, that is why the consideration
of two layers which are described by $4\times 4$ real Hamiltonians is
well justified.  Let us start again with $2\times 2$  real matrix $H$,
and choose for simplicity  the easiest form for the vector ${\bf m}({\bf
p})$. For ${\bf m}({\bf p})=  {\bf p}=(p_x,p_y)$ the Hamiltonian is
 \begin{equation} 
H=\tau_3p_x + \tau_1 p_y  ~. 
\label{P2DWaveHamiltonian}
\end{equation}
The node which we are interested in is at
$p_x=p_y=0$ and has  the topological charge (winding number)
$N_2=1$ in Eq.(\ref{Invariant}). The Dirac-type Hamiltonian
(\ref{P2DWaveHamiltonian}) and the corresponding nodes of co-dimension 2
are relevant for electrons leaving in the 2D carbon sheet known
as graphene \cite{Novoselov,Gusynin,NovoselovQHE,Falko}.

Let us now introduce two bands or layers  whose Hamiltonians  have
opposite signs:
\begin{equation} 
H_{11}=\tau_3p_x +\tau_1 p_y ~~,~~ H_{22}=-\tau_3p_x - \tau_1 p_y 
~, 
\label{P2DWaveHamiltonian-}
\end{equation}
Each Hamiltonian has a node 
at $p_x=p_y=0$. In spite of the different signs of the Hamiltonian,   the
nodes have same winding number $N_2=1$:  in the second band one has 
${\bf m}_2({\bf p})=- {\bf m}_1({\bf p})$, but
$N_2({\bf m})=N_2(-{\bf m})$ according to Eq.(\ref{Invariant}).

The Hamiltonians (\ref{P2DWaveHamiltonian}) and
(\ref{P2DWaveHamiltonian-}) can be now combined in the $4\times 4$ 
real Hamiltonian:
\begin{equation} 
H=\sigma_3 (\tau_3p_x+ \tau_1 p_y) ~,  
\label{P2DWaveHamiltonian+-}
\end{equation}
where ${\mbox{\boldmath$\sigma$}}$ matrices operate in the 2-band space. 
The Hamiltonian (\ref{P2DWaveHamiltonian+-}) has two nodes: one
is for projection $\sigma_3=1$ and another one -- for the projection
$\sigma_3=-1$. Their positions in momentum space and their topological 
charges coincide.  Let us now add the term with 
$\sigma_1$, which mixes the two bands without violation of the  time
reversal symmetry: 
\begin{equation} 
H=\sigma_3 (\tau_3p_x+ \tau_1 p_y) +\sigma_1 m~.  
\label{P2DWaveHamiltonianM}
\end{equation}
The spectrum becomes fully gapped, $E^2=p^2+m^2$, i.e. 
the two nodes annililate each other.  Since the nodes have the same
winding number $N_2$, this means that the summation law for these nodes 
is 1+1=0. Thus the zeroes of co-dimension 2 (nodal points in 2D systems
or the nodal lines in  the 3D systems) which appear in the
$4\times 4$ (and higher) real Hamiltonians are described by the 
$Z_2$-group. The discussion of the $Z_2$ nodes in
high-$T_c$ materials, polar state of $p$-wave pairing and mixed
singlet-triplet superconducting states can be found in Ref. \cite{Sato}.
 
\subsection{Gap vs splitting of nodes in bilayer materials} 
\label{GapvsSplitting}

The above example demonstrated how in the two band systems  (or in the
double layer systems) the interaction between the bands (layers) induces
the annihilation of likewise nodes and formation of the fully gapped
state. 
Experiments on the graphite film with two graphene layers
demonstrate that the spectrum of quasiparticles is essentially different
from that in a single carbon sheet \cite{NovoselovQHE}. From the detailed calculations
\cite{Falko} it follows that the gap in the spectrum emerges in the graphite bilayer at the neutrality point,
 illustrating the  rule $1+1=0$ for the $Z_2$ nodes of co-dimension 2.

Applying this to the high-T$_c$ materials
with 2, 3 or 4 cuprate layers per period, one concludes that the
interaction between the  layers can in principle induce a small gap even
in a pure $d$-wave state.  However, this does not mean that such
destruction of the Fermi lines necessarily occurs.  Instead, the
interaction between the bands (layers) can lead to splitting of
nodes, which then will occupy different positions in momentum space and
thus cannot annihilate. Which of the
two scenarios occurs -- gap formation or splitting of nodes -- depends on
the parameters of the system. Changing these parameters one can produce
the topological quantum phase transition from the fully gapped vacuum
state to the vacuum state with pairs of nodes, as we discussed for the
case of nodes with  co-dimension 3 in Sec. \ref{FermiPoints}. 
The splitting of nodes has been
observed in the bilayer cuprate Bi$_2$Sr$_2$CaCu$_2$O$_{8+\delta}$ 
 \cite{Kordyuk,Ino}.  If in the other bilayer cuprates the splitting is
not well resolved, this might be the indication of the small gap there.

\subsection{Exotic fermions and reentrant violation of `special
relativity' in bilayer graphene} 
\label{ReentrantViolation}

There still can be some discrete symmetry which forbids the 
annihilation of nodes of co-dimension 2, even if the nodes
are not separated in momentum space. This is the
symmetry between the two layers which forbids the rule $1+1=0$. For
example, if the  Hamiltonian still anti-commutes with some matrix, say,
with
$\tau_2$-matrix, there is a generalization of the integer valued invariant
in Eq.(\ref{Invariant}) to the $2n\times 2n$ real
Hamiltonian (see also \cite{WenZee}):
 \begin{equation} 
N_2=- {1\over 4\pi i} ~{\rm tr} ~\oint dl ~ \tau_2 H^{-1}\nabla_l H~. 
\label{Invariant2}
\end{equation}
 Since the summation law for this $N_2$ charge is 1+1=2, the nodes
with $N_2=1$ present at each layer do not annihilate each other if the
interaction term preserves the symmetry. In this case the spectrum of
the bilayer system remains gapless.

\begin{figure}[t]
\centerline{\includegraphics[width=1.00\linewidth]{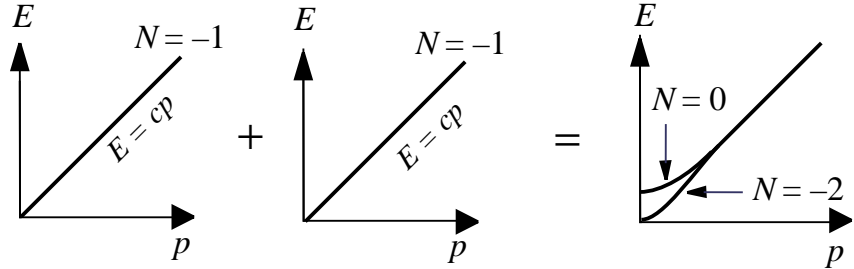}}
%\centerline{\epsfxsize=0.90\textwidth\epsfbox{DresdenReentrant.eps}}
\medskip
\caption{Reentrant violation of special relativity in 2D and 3D systems.
Close to the topologically stable zero, single fermion has a relativistic
spectrum. $N$ stands for the topological charge
$N_2$ in 2D systems and  the topological charge $N_3$ for 3D systems.
In case of several relativistic fermions with the same topological charge,
the interaction between the fermions may lead to the rearrangement of the
topological charges between the fermions, 
$(-1)+(-1)\rightarrow (-2) +0$. This means that mixing between two
left-handed neutrinos with charges $N_3=-1$ each results in one exotic
non-relativistic neutrino with
$N_3=-2$ and one massive neutrino with $N_3=0$. The same occurs for the
fermions in the bilayer graphene, $1+1\rightarrow 2 +0$: due to
interaction between the layers, two Dirac fermions (one per each layer) with
spectrum
$E=cp$ transform to  one exotic non-relativistic fermion with $N_2=2$ and
quadratic spectrum
$p^2/2m$ and one fully gapped fermion with $N_2=0$.}
\label{DresdenReentranFig}
\end{figure}

Let us now consider the gapless spectrum in such bilayer material.
We start again with the Hamiltonian in  Eq.(\ref{P2DWaveHamiltonian+-}),
which describes gapless Dirac quasiparticles living in two independent
layers, and add the interaction between them which does not violate the 
$\tau_2$-symmetry:
 \begin{equation} 
H=\sigma_3 (\tau_3p_x+ \tau_1 p_y) +m (\tau_1\sigma_1- \tau_3 \sigma_2) ~.
\label{CarbonBilayer}
\end{equation}
The energy spectrum becomes
 \begin{equation}
E_+= \pm \left(\sqrt{m^2  +p ^2}+m\right) ~~,~~ E_-= \pm \left( 
\sqrt{m^2  +p ^2}-m\right) ~ .
\label{CarbonBilayerSpectrum}
\end{equation}
Without interaction, i.e. at $m=0$, the quasiparticles represent two Dirac
fermions with the topological charges $N_2=1$ each.   Since the Hamiltonian
(\ref{CarbonBilayer})  anti-commutes with the
$\tau_2$-matrix, the total  topological charge $N_2$ must be conserved
even at $m\neq 0$. Thus the total charge for quasiparticles 
must be $N_2= 2$. However it is now distributed between the branches of
quasiparticle spectrum in the following manner (Fig.
\ref{DresdenReentranFig}). For $m>0$, the quasiparticles with energy $E_+$ 
acquire the trivial topological charge
$N_2=0$,  that is why their spectrum becomes fully gapped: $E_+(p\ll
m)\approx\pm 2m$. The quasiparticles with energy  $E_-$   have the rest
nonzero topological charge $N_2= 2$, and thus they must be gapless. The
energy spectrum of these gapless fermions with $N_2= 2$ is exotic: at
$p\ll m$ the spectrum becomes that of classical particles  with
positive and negative masses, $E_-\approx \pm p^2/2m$;  in the region
$p\gg m$ it is relativistic $E\approx
\pm p$; and finally  the relativistic invariance is violated again at high
$p$ of order of inverse inter-atomic distance.  When the parameter $m$
crosses zero, the quantum phase transition occurs.
    
It is important  that the exotic branch with $N_2= 2$ contains only single
fermionic species, i.e. it cannot split into two fermions  with $N_2= 1$
each. That is why the quadratic law for the spectrum of exotic fermions is
generic, provided that the proper symmetry of the Hamiltonian is obeyed. 
The same spectrum (\ref{CarbonBilayerSpectrum}) takes place for
quasiparticles in the carbon film consisting of  two graphene sheets: it
occurs in some range of parameters of the system where terms in the
Hamiltonian, which violate the $\tau_2$-symmetry and induce the gap in the
spectrum, are small and can be neglected
\cite{Falko}. Exotic fermions with parabolic
spectrum lead to the unconventional quantum Hall effect
\cite{Falko}, which has been observed in the bilayer graphene  
\cite{NovoselovQHE}. 
  
All this shows that the stability of and the summation law for the nodal 
lines  depend on the type of discrete symmetry which protects the topological 
stability.   The integer valued topological invariants protected 
  by discrete or continuous symmetry were discussed in Chapter 12 of the 
book \cite{VolovikBook}.

If the symmetry is obeyed we have the following situation. Fermions with
the elementary topological charge, $N_2=\pm 1$, are necessarily
relativistic in the low-energy corner,  according to the
Atiyah-Bott-Shapiro construction.  However, even a very small interaction
between two species with $N_2=+1$ each may produce the exotic fermions,
which are classical. In this scenario the Lorentz invariance is violated
both at very high and at very low energies, therefore the term `reentrant
violation of special relativity'.

\subsection{Reentrant violation of special relativity in 3D systems} 
\label{ReentrantViolation3D}

Similar reentrant violation of Lorentz invariance in the 3D vacua may
occur for the Fermi points of co-dimension 3 described by the topological
charge $N_3$ \cite{VolovikReentrant,VolovikBook}. Let us suppose  that the
Standard Model is an effective theory, and  that the right-handed
neutrinos are absent in this theory. The left-handed neutrino, which has
$N_3=-1$, is necessarily massless. Its spectrum is necessarily
relativistic in the low-energy corner, and thus the Lorentz invariance
emerges at low energies. However, according to the general rules, if 
several species of the left-handed neutrino are present, so that 
$|N_3|>1$, the necessary violation of the  Lorentz invariance at high energy
must induce the violation of the  Lorentz invariance at very low energy. Let
us consider the case of two flavors of the left-handed
neutrino -- electron and muon neutrinos with
$N_3=-1$  each  in Fig.
\ref{DresdenReentranFig} ({\it left}). If the Standard Model is an effective
theory, the high-energy cut-off induces mixing between the two flavors. The
mixing  is rather small, since it contains the high-energy cut-off in the
denominator. Such a mixing typically  leads to the
rearrangement of the topological charges:
$(-1)+(-1)\rightarrow (-2) +0$. This means that the two
chiral neutrinos transform into one exotic non-relativistic neutrino with
$N_3=-2$ and one massive neutrino with $N_3=0$ in  Fig.
\ref{DresdenReentranFig} ({\it right}). In the particular model
discussed in Refs.
\cite{VolovikReentrant,VolovikBook}, the corresponding spectrum of two
neutrino flavors is
 \begin{equation}
E_\pm^2=  p_z^2+\left( \sqrt{m^2  +p_x^2+p_y^2} \pm m\right)^2 ~ .
\label{SMSpectrum}
\end{equation}
At $p_z=0$, this 3D spectrum transforms to the 2D spectrum in
Eq.(\ref{CarbonBilayerSpectrum}). 
 The magnitude $m$ of the splitting of the neutrino spectrum has been
discussed in Ref.  \cite{KlinkhamerReentrant}. 
Some speculations on the possible consequences of the reentrant violation
of special relativity are discussed in Ref. 
\cite{Consoli}.

\subsection{Quantum phase transition in high-T$_c$ superconductor} 
\label{TransitionNodalSuperconductor}

\begin{figure}[t]
\centerline{\includegraphics[width=0.80\linewidth]{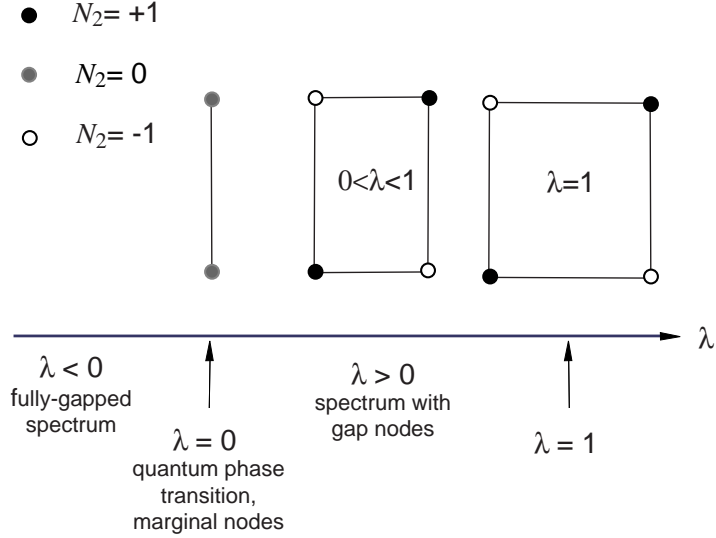}}
%\centerline{\epsfxsize=0.70\textwidth\epsfbox{Quadrate.eps}}
\medskip
\caption{Quantum phase transition by change of anisotropy parameter
$\lambda$ in Eq. (\ref{dWaveHamiltonian}) for superconductors in the $d+s$
state.  At $\lambda>0$ the 2D spectrum has 4 nodes: two with
topological charge $N_2=+1$ ({\it filled circles}) and 
two with
topological charge $N_2=-1$ ({\it open circles}). At $\lambda=0$, points
with opposite charges merge forming two marginal nodes with 
$N_2=0$ ({\it grey circles}). The marginal (topologically trivial) nodes
disappear at $\lambda<0$ leaving the fully gapped vacuum }
\label{QuantPTransitionFig4}
\end{figure}

 Let us return to the $2\times 2$ real Hamiltonian 
(\ref{dWaveHamiltonian}) and consider what happens with gap nodes when
one changes the asymmetry parameter
$\lambda$.  When
$\lambda$ crosses zero  there is a quantum phase transition at which
nodes in the spectrum annihilate each other and then the fully gapped
spectrum develops [Fig. \ref{QuantPTransitionFig4}]. Note that there is no
symmetry change across the phase transtion.

The similar quantum phase transition from gapless to gapped state without
change of symmetry also occurs when
$\mu$ crosses zero.  This scenario can be realized in the BEC--BCS
crossover region, see
\cite{Botelho,Borkowski,Duncan}. 

The presence of the gap nodes in high-$T_c$ superconductors is indicated
by the measurement of the field dependence of electronic
specific heat $C$ at low temperatures. If the superconducting state is
fully gapped, then $C \propto H$; while
if there are point nodes in 2D momentum space  then the heat capacity
is nonlnear, $C \propto \sqrt{H}$ \cite{Volovik1993}. An unusual behavior
of $C$ in high-$T_c$ cuprate  Pr$_{2-x}$Ce$_{x}$CuO$_{4-\delta}$ has been
reported in Ref. \cite{Balci}.  It was found that the field dependence of
electronic specific heat is linear at $T=2$K, and non-linear at $T\geq$3K.
If so, this behavior could be identified with the quantum phase transition
from gapped to gapless state, which is smeared due to finite temperature. 
However, the more accurate measurements have not confirmed the change
of the regime:  the nonlinear behavior $C\propto \sqrt{H}$ continues below
$T=2$K \cite{Green}.

\section{Topological transitions in fully gapped systems} 
\label{PlateauTransitions} 

\subsection{Skyrmion in 2-dimensional momentum space} 

The fully gapped ground states (vacua) in 2D systems or in quasi-2D thin
films, though they do not have zeroes in the energy spectrum, can also be
topologically non-trivial. They are characterized by the invariant 
which is the dimensional reduction of the topological invariant for the
Fermi point in Eq.(\ref{TopInvariant}) \cite{Ishikawa,VolovikYakovenko}:  
\begin{equation} 
\tilde N_3 = {1\over{24\pi^2}}e_{\mu\nu\lambda}~
{\bf tr}\int   d^2pd\omega
~ G\partial_{p_\mu} G^{-1}
G\partial_{p_\nu} G^{-1} G\partial_{p_\lambda}  G^{-1}~.
\label{3DTopInvariant}
\end{equation} 
For the fully gapped vacuum, there is no singularity in the Green's
function, and thus the integral over the entire 3-momentum space
$p_\mu=(\omega,p_x,p_y)$ is well determined. If a crystalline system is
considered the integration over $(p_x,p_y)$ is bounded by the Brillouin
zone. 

\begin{figure}[t]
\centerline{\includegraphics[width=\linewidth]{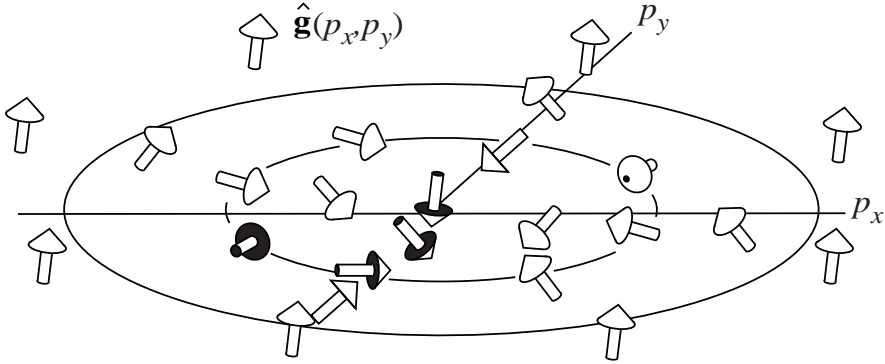}}
%\centerline{\epsfxsize=0.70\textwidth\epsfbox{PSpaceSkyrmions.eps}}
\medskip
\caption{Skyrmion in ${\bf p}$-space with momentum space topological
charge 
$\tilde N_3=-1$. It describes topologically non-trivial vacua in 2+1
systems with a fully non-singular Green's function}
\label{PSpaceSkyrmionsFig}
\end{figure}

An example is
provided by the 2D version of the Hamiltonian  (\ref{BogoliubovNambuH})
with
$\hat{\bf l}=\hat{\bf z}$, $\hat{\bf e}_1=\hat{\bf x}$,  $\hat{\bf
e}_2=\hat{\bf y}$. Since for 2D case one has
$p^2=p_x^2+p_y^2$,  the quasiparticle energy (\ref{BogoliubovNambuE})
\begin{equation}
E^2 ({\bf p}) = \left({p_x^2+p_y^2\over 2m}-\mu\right)^2+\,
          c^2(p_x^2+p_y^2) ~ 
\label{BogoliubovNambuE2D}
\end{equation}
 is nowhere zero except for $\mu=0$. 
The Hamiltonian (\ref{BogoliubovNambuH}) can be written in terms of 
the three-dimensional vector ${\bf g}(p_x,p_y)$:
\begin{equation}
{\cal H}=\tau_i g_i({\bf p})~~,~~g_3 ={p_x^2+p_y^2\over
2m}-\mu~,~g_1=  c p_x~,~g_2=  -c p_y  ~~.
\label{2DFullyGapped}
\end{equation}
For $\mu>0$ the distribution of the unit vector $\hat{\bf
g}(p_x,p_y)= {\bf
g}/|{\bf
g}|$ in the momentum space has the same structure as the skyrmion in real
space (see Fig.
\ref{PSpaceSkyrmionsFig}). The topological invariant  for
this momentum-space skyrmion is given by Eq.(\ref{3DTopInvariant})  which
can be rewritten in terms of the unit vector  $\hat{\bf g}(p_x,p_y)$:
\begin{equation}
\tilde N_3= {1\over 4\pi}\int dp_xdp_y~\hat{\bf g}\cdot
\left({\partial \hat{\bf g}\over\partial {p_x}} \times {\partial \hat{\bf
g}\over\partial {p_y}}\right)~.
\label{2DInvariant}
\end{equation}
Since at infinity the unit vector field $\hat{\bf g}$ has
the same value, $\hat{\bf g}_{p \rightarrow
\infty}
\rightarrow (0,0,1)$, the 2-momentum space 
$(p_x,p_y)$ becomes isomoprhic to the compact
$S^2$ sphere. The function $\hat{\bf g}({\bf p})$ realizes the
mapping of
this $S^2$ sphere to the $S^2$ sphere of the unit vector  $\hat{\bf g}$ 
with winding number $\tilde N_3$. For $\mu>0$ one has  $\tilde N_3=-1$ 
and for $\mu<0$ one has  $\tilde N_3=0$.

\subsection{Quantization of physical  parameters}
\label{QuantizationPphysicalParameters}

The topological charge  $\tilde N_3$ and other similar topological
charges in 2+1 systems give rise  to quantization parameters.  In
particular, they are responsible for quantization of Hall and spin-Hall
conductivities, which occurs without applied magnetic field (the
so-called intrinsic or anomalous quantum Hall and spin quantum Hall 
effects). There are actually 4 responses of currents to transverse forces
which are quantized under appropriate conditions. These are: (i)
quantized response of the mass current (or electric current in
electrically charged systems) to transverse gradient of chemical potential
$\nabla
\mu$ (transverse electric field ${\bf E}$); (ii) quantized response of 
the mass current (electric current) to transverse gradient of magnetic
field interacting with Pauli spins; (iii) quantized response of the spin
current to transverse gradient of magnetic field; and (iv) quantized
response of the spin current to transverse gradient of chemical potential
(transverse electric field)
\cite{SQHE}. 

\subsubsection{Chern-Simons term and ${\bf p}$-space topology}

All these responses can be described using the generalized Chern-Simons 
term which mixes different gauge fields (see Eq.(21.20) in Ref.  
\cite{VolovikBook}):
\begin{equation}
F_{\rm CS}\{{\bf A}_Y\}= {1\over 16\pi} N_{IJ} e_{\mu\nu\lambda}\int
d^2xdt A_\mu^IF^J_{\nu\lambda}  ~~.
\label{ChernSimons}
\end{equation}
Here $ A_\mu^I$ is the set of the real or auxiliary (fictituous) gauge
fields.   In
electrically neutral systems, instead of the gauge field
$A_\mu$ one introduces the  auxiliary $U(1)$ field, so that the current
is given by variation of the action with respect to $A_\mu$:  $\delta
S/\delta A_\mu=J^\mu$. The  auxiliary
$SU(2)$ gauge field $A_{\mu}^i$ is convenient for the description of the
spin-Hall effect, since the variation of the action with respect to
$A_\mu^a$ gives the spin current: $\delta S/\delta A_{\mu}^i=J^\mu_i$.
Some components of the field $A_{\mu a}$ are physical, being represented
by the real physical quantities which couple to the fermionic charges.
Example is provided by the external magnetic field in neutral system,
which play the role of $A_{0}^i$ (see Sec. 21.2 in Ref.
\cite{VolovikBook}).  After the current is calculated the values of the
auxiliary fields are fixed. The latest discussion of the mixed
Chern-Simons term can be found in Ref.  \cite{MutualCS}. For the related
phenomenon of axial anomaly, the mixed action in terms of different
(real and fictituous) gauge fields has been introduced in Ref.
\cite{Zhitnitsky}.

The important fact is that the matrix $N_{IJ}$ of the prefactors
in the Chern-Simons  action is expressed in terms of the momentum-space
topological invariants:
\begin{equation} 
N_{IJ} = {1\over{24\pi^2}}e_{\mu\nu\lambda}~
{\bf tr}~ Q_IQ_J\int   d^2pd\omega
~ G\partial_{p_\mu} G^{-1}
G\partial_{p_\nu} G^{-1} G\partial_{p_\lambda}  G^{-1}~,
\label{ProtectedTopInvariant}
\end{equation} 
where $Q_I$ is the fermionic charge interacting with the gauge
field 
$ A_\mu^I$ (in case of several fermionic species,  $Q_I$ is a
matrix in the space of species).

\subsubsection{Intrinsic spin quantum Hall effect}

To obtain, for example, the response of
the spin current 
$j_z^i$ to the electric field $E_i$, one must consider two
fermionic charges: the electric charge
$Q_1=e$ interacting with $U(1)$ gauge field, and the spin along
$z$ as another charge,
$Q_2=s_z=\hbar\sigma_z/2$, which interacts with the fictituous $SU(2)$
field $A_{\mu}^z$. This gives the quantized spin current response to the
electric field
$j_z^i=e^{ij}\sigma_{\rm spin-Hall}E_j$, where 
$\sigma_{\rm spin-Hall}=(e\hbar/8\pi)N$ and $N$ is integer:
\begin{equation} 
N  = {1\over{24\pi^2}}e_{\mu\nu\lambda}~
{\bf tr}~ \sigma_z \int   d^2pd\omega
~ G\partial_{p_\mu} G^{-1}
G\partial_{p_\nu} G^{-1} G\partial_{p_\lambda}  G^{-1}~.
\label{ProtectedTopInvariant2}
\end{equation} 
 Quantization of  the spin-Hall
conductivity in the commensurate lattice of vortices can be found in Ref.
\cite{Vafek}.

The above consideration is applicable, when the momentum (or
quasi-momentum in solids) is the well defined quantity,  otherwise (for
example, in the presence of impurities) one cannot construct the
invariant in terms of the Green's function
$G({\bf p},\omega)$.  However, it is not excluded that  in some cases the
perturbative introduction of impurities does not change  the prefactor
$N_{IJ}$ in the Chern-Simons term (\ref{ChernSimons}) and thus does not
influence the quantization: this occurs if there is no spectral flow
under the adiabatic introduction of impurities. In this case the
quantization is determined by the reference system -- the fully gapped
system from which the considered system can be obtained by the continuous
deformation without the spectral flow (analogous phenomenon for the
angular momentum paradox in $^3$He-A was discussed in
\cite{AngularMomentumPar}). The most recent review paper  on the spin
current can be found in
\cite{Rashba}.

\subsubsection{Momentum space topology and Hall effect in 3D systems}

The momentum space topology is also important for the Hall effect  in
some 3+1 systems. The contribution of  Fermi points to the intrinsic Hall
effect is discussed in the Appendix of Ref. \cite{SplittingPreprint}. 
For metals with Fermi surfaces having the global topological charge $N_3$
(see Sec. \ref{FermiSurfaceGlobalCharge}) the anomalous Hall effect is
caused by the Berry curvature on the Fermi surface  \cite{Haldane}. The
magnitude of the Hall conductivity is related to the volume of the Fermi
surface in a similar way as  the number of particles and the volume of
the Fermi surface are connected by the Luttinger theorem \cite{Haldane}.
Another ``partner'' of the Luttinger theorem  emerges for the
Hall effect in superconductors, where topology
enters via the spectral flow of fermion zero modes in the cores of
topological defects -- Abrikosov vortices
\cite{EffectBandStructure}. 

\subsection{Quantum phase  transitions} 
\subsubsection{Plateau transitions}

\begin{figure}[t]
\centerline{\includegraphics[width=\linewidth]{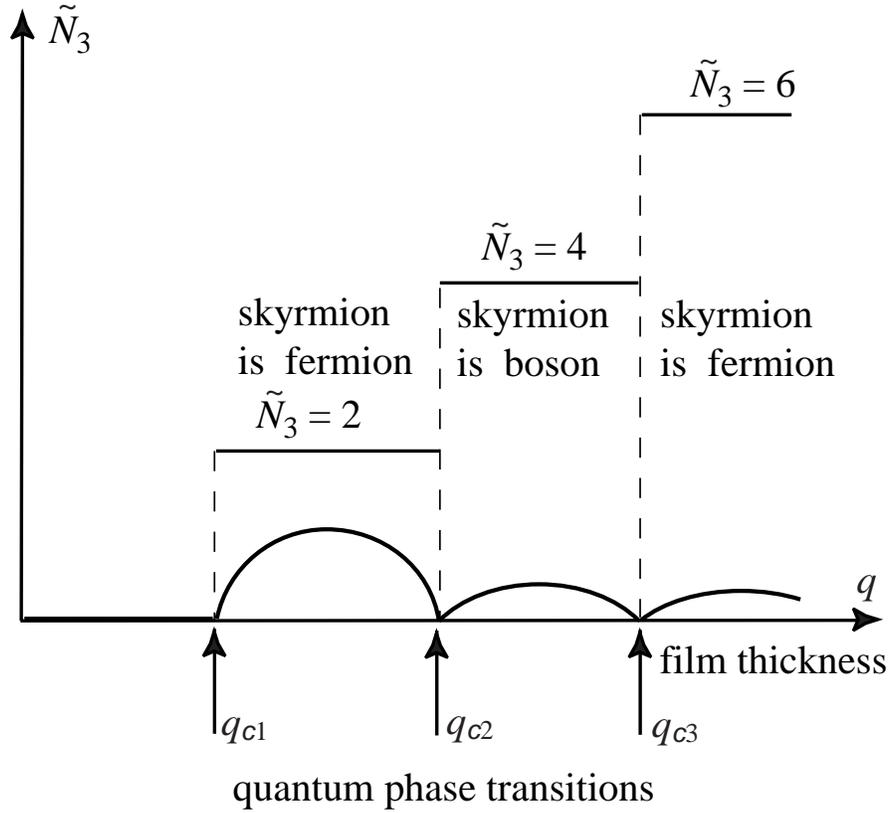}}
%\centerline{\epsfxsize=0.70\textwidth\epsfbox{FermionBoson.eps}} 
\medskip
\caption{Quantum phase transitions occurring when one increases  
the  thickness $q$ of the $^3$He-A film. The transitions at $q=q_{c2}$ 
and
$q=q_{c3}$ are plateau-plateau transitions between vacua with different 
values of integer topological invariant
$\tilde N_3$ in Eq.(\ref{3DTopInvariant}). At these transitions  the
quantum statistics of real-space skyrmions living in thin films changes.
The change in the quasiparticle spectrum across the transitions may be seen
from the minimum value of the
quasiparticle energy, ${\rm min}_{\bf p}E({\bf p})$, at given
$q$ ({\it thick lines}). The transitions at
$q=q_{c2}$ and
$q=q_{c3}$ between the fully gapped states occur through the gapless
states. At
$q=q_{c1}$ the transition is between gapless and fully gapped states
 }
\label{FermionBosonFig}
\end{figure}

The integer topological invariant $\tilde N_3$ of the ground state cannot 
follow  the continuous parameters of the system. That is why when one
changes such a parameter, for example, the chemical potential in the
model (\ref{2DFullyGapped}),  one obtains the quantum phase transition at
$\mu=0$ at which  $\tilde N_3$ jumps from $0$ to $-1$. The film thickness
is another relevant parameter.  In the film with finite thickness the
matrix of Green's function  acquires indices of the levels of transverse
quantization. If one increases the thickness of the film,  one finds a
set of quantum phase transitions between vacua with different integer
values of the invariant [Fig.
\ref{FermionBosonFig}], and thus between the plateaus in Hall or spin-Hall
conductivity. 

The abrupt change of the topological charge 
cannot occur adiabatically, that is why  at the points of
quantum transitions fermionic quasiparticles become gapless.

 \subsubsection{Topological zero modes and edge states}

\begin{figure}[t]
\centerline{\includegraphics[width=0.80\linewidth]{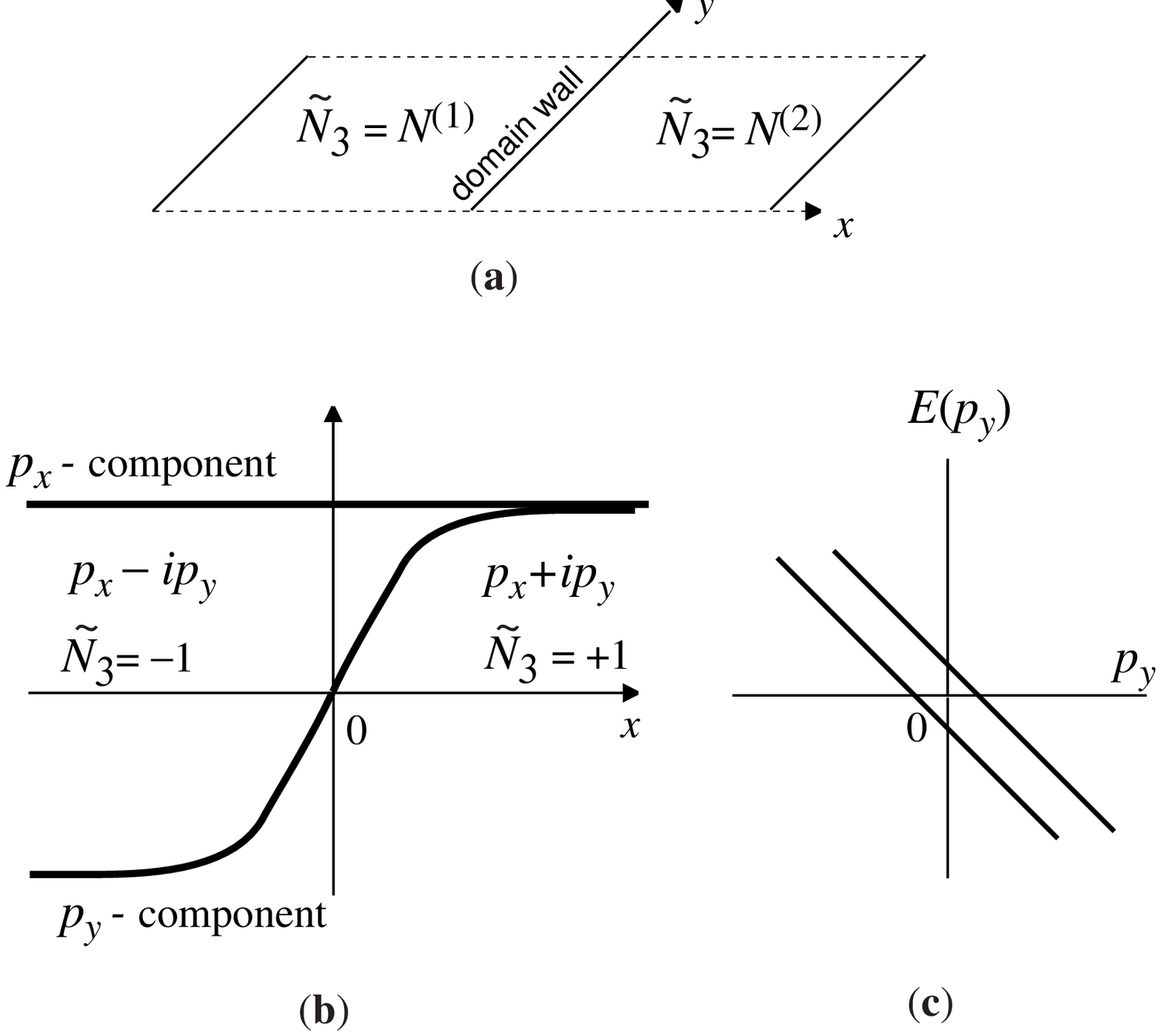}}
%\centerline{\epsfxsize=0.70\textwidth\epsfbox{EdgeStates.eps}}
\medskip
\caption{({\bf a}) Domain wall between two 2+1 vacua with different
topological charges $\tilde N_3$. ({\bf b}) Structure of the phase
boundary between vacua with charges $\tilde N_3=\pm 1$
in Eq.(\ref{EdgeStates}). The prefactor in front of
$p_y$ changes sign at $x=0$, which leads to the change of sign of the
topological charge in Eq.(\ref{3DTopInvariant}). ({\bf c})
Fermion zero modes -- anomalous branches of fermions living at the
interface whose spectrum crosses zero energy level. The number of
anomalous branches is determined by the difference of the
topological charges
$\tilde N_3$ across the wall}
\label{EdgeState}
\end{figure}

If two vacua with different  $\tilde N_3$ coexist in space
[Fig. \ref{EdgeState}({\bf a})], the phase boundary between them must also
contain gapless fermions. This is an example of the so-called fermion zero
modes living on different topological objects such as 3D monopole, 2D
soliton wall, and 1D vortex/string (see Ref.
\cite{JackiwRossi} and references therein). The number of the gapless
fermion zero modes obeys the index theorem: in our case the number of the
$1+1$ fermions living at the phase boundary is determined by the
difference of the topological charges of the two vacua,
$\tilde N_3^{(1)} -  
\tilde N_3^{(2)}$ (see Chap. 22 in Ref. \cite{VolovikBook}).

The boundary of the condensed matter system can be considered as the phase
boundary between the state with nonzero $\tilde N_3$ and the state with 
$\tilde N_3=0$. The corresponding fermion zero modes are 
the edge states well known in physics of the QHE.

Example of the phase boundary between two vacua with $\tilde N_3=\pm 1$ is
shown in Fig. \ref{EdgeState}({\bf b}) for the $p_x+ip_y$ superfluids and
superconductors. Here the $p_y$ component of the order parameter changes
sign across the wall.   The simplest structure of such boundary is given by
Hamiltonian
\begin{equation}   
H=  
\left(\matrix{ \frac{p^2}{2m}  -\mu
&c\left(p_x+ ip_y \tanh \frac{x}{\xi}\right) \cr
   c\left(p_x- ip_y \tanh \frac{x}{\xi}\right)
&- \frac{p^2}{2m}  +\mu \cr }\right)~.
\label{EdgeStates}
\end{equation}
Let us first consider fermions in semiclassical approach, when the
coordinates $x$ and $p_x$ are independent. At $x=0$ the time reversal
symmetry is restored, and the spectrum becomes gapless.   At $x=0$ there
are two zeroes of co-dimension 2 at points $p_x=0$ and $p_y=\pm p_F$. They
are similar to zeroes discussed in Sec. \ref{Z2lines}. These zeroes are
marginal, and disappear at $x\neq 0$ where the time reversal symmetry is
violated.  The topological charge is well defined only at $x\neq 0$.
 When $x$ crosses zero, the topological charge in
Eq.(\ref{3DTopInvariant}) changes sign.

In the quantum mechanical description, $x$ and $p_x$ do not commute. The
quantum-mechanical spectrum $E(p_y)$ contains fermion zero modes --
branches of spectrum which cross zero. According to the
index theorem there are two anomalous branches in Fig.
\ref{EdgeState}({\bf c}).  

The index theorem together with the
connection between the topological charge and quantization of physical
parameters discussed in Sec.
\ref{QuantizationPphysicalParameters} implies that the quantization of
Hall and/or spin Hall conductance is determined by the number of edge
states in accordance with Refs. \cite{WenStone}.
The detailed discussion of the edge modes in $p_x+ip_y$ superfluids and
superconductors and their contribution to the effective action can be
found in Ref. \cite{Stonepxpy}. These edge modes are Majorana fermions.
Recent discussion of the edge states and intrinsic quantum
spin-Hall effect in graphene is in Ref. \cite{Sengupta}.

 \subsubsection{``Higgs'' transition in ${\bf p}$-space}
 
\begin{figure}[t]
\centerline{\includegraphics[width=0.5\linewidth]{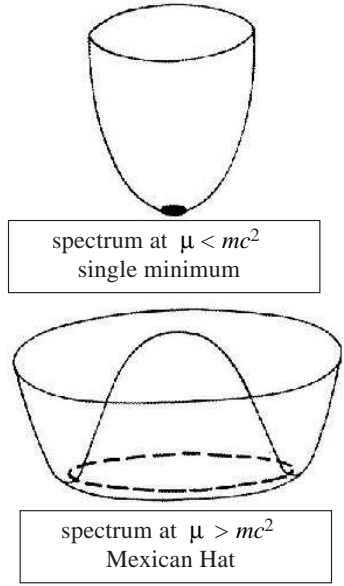}}
%\centerline{\epsfxsize=0.60\textwidth\epsfbox{TopolofyOfMinima.eps}}
\medskip
\caption{
``Higgs'' transition in momentum space.
}
\label{TopolofyOfMinimaFig}
\end{figure}

Note that the energy spectrum in Eq.(\ref{BogoliubovNambuE2D}) experiences
an analog of the Higgs phase transition at $\mu=mc^2$
[Fig. \ref{TopolofyOfMinimaFig}]: if $\mu<mc^2$ the
quasiparticle energy has a single minimum at $p=0$, while at $\mu>mc^2$
the minimum is at the circumference with radius $p_0=\sqrt{2m(\mu
-mc^2)}$. There is no symmetry breaking at this transition, since the
vacuum state has the same rotational symmetry above and below
the transition, while the asymptotic behavior of the thermodynamic 
quantities ($\propto T^n\exp{(-E_{\rm min}/T})$)
 experiences discontinuity across the transition: the power $n$ changes. 
That is why the point $\mu=mc^2$ marks the quantum phase transition, at
which the topology of the minima of the energy spectrum changes. 
 
 However, this transition does not belong to the class of
transitions which we discuss in the present review, since the topological
invariant  of the ground state $\tilde N_3$ does not change across this
transition and thus  at the transition point  $\mu=mc^2$ the spectrum 
remains fully gapped. Moreover, such a transition does not depend on
dimension of space-time
 and occurs in 3+1 systems  as well. Example is provided by the $s$-wave 
superconductor
 or $s$-wave Fermi superfluid, whose spectrum in Eq.(\ref{sWave}) 
experiences the same Higgs-like transition at $\mu=0$, i.e. in the
BSC--BEC crossover region. 

\subsection{Quantum phase transition in 1D quantum Ising model}

The momentum-space topology is applicable not only to fermionic systems, 
but to any system which can be expressed in terms of auxiliary fermions.

\subsubsection{Fermionization and topological invariant} 

Example is provided by the 1-dimensional quantum Ising
model where the topological quantum phase transition between the fully 
gapped vacua can be described in terms of the invariants for the
fermionic Green's function. The original  Hamiltonian of this 1D chain of
spins is:
\begin{equation}
H=-J\sum_{n=1}^N \left( h\sigma_n^x +\sigma_n^z\sigma_{n+1}^z \right)  ~~,
\label{Isings}
\end{equation}
where $\sigma^x$ and $\sigma^z$ are Pauli matrices, and $h$  is the
parameter describing the external magnetic field. After the standard
Jordan-Wigner transformation this system can be represented in terms of
the non-interacting fermions with the following Hamiltonian in the
continuous
$N\rightarrow
\infty$ limit (see Ref.
\cite{Dziarmaga} and references therein):
\begin{equation}
H= 2J\left(h -\cos(pa)\right)\tau_3 +  
2J\sin(pa) \tau_1~~,~~
-\frac{\pi}{a} < p<\frac{\pi}{a} ~~.
\label{IsingFermionsHam}
\end{equation}
It is periodic in the one-dimensional momentum space $p$ with period 
$2\pi/a$  where $a$ is the lattice spacing.
The integer
valued topological invariant here is the same as in Eq.
(\ref{Invariant2}) but now the integration is along the closed path  in
$p$-space, i.e. from $0$ to $2\pi/a$: 
\begin{equation} 
\tilde N_2=- {1\over 4\pi i} ~{\rm tr} ~\oint dp ~ \tau_2 H^{-1}\nabla_p
H~. 
\label{Invariant2Instanton}
\end{equation}
This invariant can be represented in terms of the Green's  function
\begin{equation}
G^{-1}=ig_z -  g_x \tau_3 +   g_y \tau_1~,
\label{IsingFermions2}
\end{equation}
where for the particular case of the model (\ref{IsingFermionsHam}),  the
components of the 3D vector
${\bf g}(p,\omega)$ are: 
\begin{equation}
g_x(p,\omega)=2J\left(h -\cos(pa)\right)~~,~~
g_y(p,\omega)=2J\sin(pa)~~,~~  g_z(p,\omega)=\omega ~.
\label{Components}
\end{equation}
Then the invariant (\ref{Invariant2Instanton}) becomes:
\begin{equation}
\tilde N_2= {1\over 4\pi}\int_{-\pi/a}^{\pi/a} dp\int_{-\infty}^{\infty}  
d\omega~\hat{\bf
g}\cdot
\left({\partial \hat{\bf g}\over\partial {p}} \times {\partial \hat{\bf
g}\over\partial {\omega}}\right)~.
\label{2DInvariantIsing}
\end{equation}
The invariant is well defined for the fully gapped states,  when ${\bf
g}\neq 0$ and thus the unit vector $\hat{\bf g}= {\bf g}/|{\bf g}|$ has
no singularity.   In the model under discussion, one has for $h\neq 1$:
\begin{equation}
\tilde N_2(h<1)=1~~,~~\tilde N_2(h>1)=0~~.
\label{InvariantIsingSkyrmion}
\end{equation}

\subsubsection{Instanton in $(p,\omega)$-space} 

\begin{figure}[t]
\centerline{\includegraphics[width=0.8\linewidth]{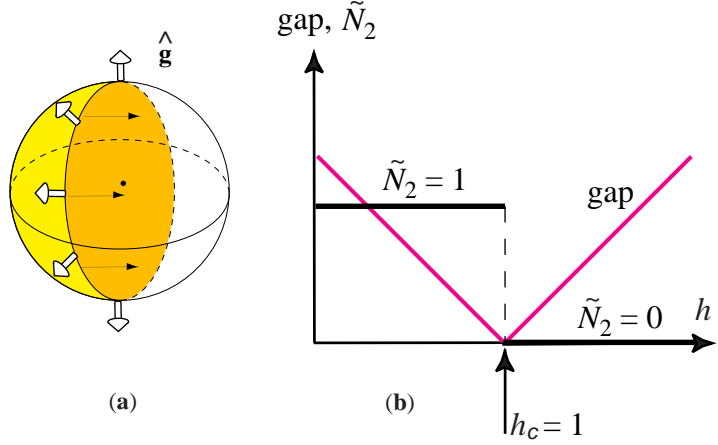}}
%\centerline{\epsfxsize=0.70\textwidth\epsfbox{P-SpaceInstanton.eps}} 
\medskip
\caption{({\bf a}) Illustration of the topological invariant $\tilde
N_2=1$ for `instanton' in momentum space for $h=0$. According to
Eq.(\ref{precession}) one has the domain wall in $z=\omega/2J$ space
across which the direction of the vector ${\bf g}$ changes from $\hat{\bf
z}$ at $z=\infty$ to  $-\hat{\bf
z}$ at  at $z=-\infty$.  The structure is periodic in $p$ and thus is 
precessing in `time'
$t=p$ ({\it black arrows}). During one period of precession $\Delta
t=2\pi/a$ the  unit vector
$\hat{\bf g}(t,z)$
 sweeps the whole unit sphere giving $\tilde
N_2=1$ in Eq.(\ref{2DInvariantIsing}). ({\bf b}) At the transition point 
$h_c=1$ the gap in the energy spectrum of fermions vanishes, because the
transition between two vacuua with different topological charge cannot
occur adiabatically }
\label{P-SpaceInstantonFig}
\end{figure}

The state with $\tilde N_2=1$ is
the ``instanton'' in the $(\omega,p)$-space, which is similar to the  
skyrmion in
$(p_x,p_y)$-space in Fig.
\ref{PSpaceSkyrmionsFig}. 
The real space-time counterpart of such instanton  can
be found in Refs.
\cite{Instanton}. It describes the periodic phase slip process occurring
in  superfluid $^3$He-A  \cite{InstantonExp}.
 In the model, the topological structure of the instanton at $h<1$ can be
easily revealed for
$h=0$. Introducing ``space-time'' coordinates  $t=p$ and $z=\omega/2J$
one obtains that the unit vector $\hat{\bf g}$ precesses sweeping the
whole unit sphere during one period $\Delta t=2\pi/a$ [Fig.
\ref{P-SpaceInstantonFig}]:
\begin{equation}
\hat{\bf g}(z,t)=\hat{\bf z}\cos \theta(z) + \sin \theta(z)
\left(\hat{\bf x}\cos(at)+\hat{\bf y}\sin(at)\right)
~~,~~\cot\theta(z)=z~.
\label{precession}
\end{equation}
This state can be referred to as `ferromagnetic', since in terms of 
spins it is the quantum superposition  of two ferromagnetic states with
opposite magnetization. 

At $h>1$, i.e. in the `paramagnetic' phase,   the momentum-space
topology is trivial, $\tilde N_2(h>1)=0$. The transition at $h=1$ at which 
the topological  charge $\tilde N_2$ of the ground state changes  is the
quantum phase transition, it only occurs at $T=0$. 

\subsubsection{Phase diagram for anisotropic XY-chain}

\begin{figure}[t]
\centerline{\includegraphics[width=0.5\linewidth]{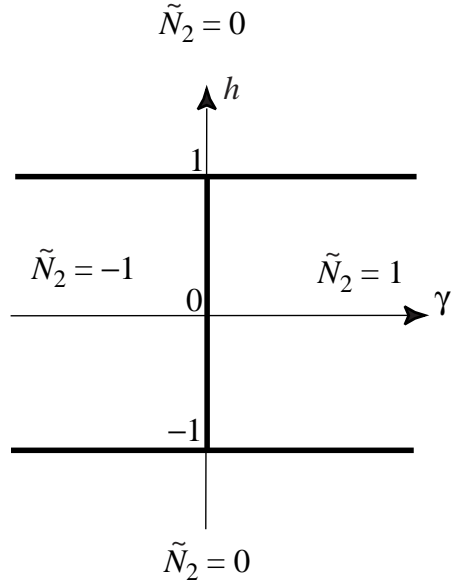}}
%\centerline{\epsfxsize=0.70\textwidth\epsfbox{AnisotropicXY.eps}} 
\medskip
\caption{Phase diagram for anisotropic XY-chain in Eq.(\ref{AnisotropicXY}) in the plane 
($\gamma,h)$. The regions with different topological charge  $\tilde N_2$
are separated by the lines of topological quantum phase transitions
({\it thick lines})}
\label{AnisotropicXYFig}
\end{figure}

The phase diagram for the extension of the Ising model  to the case of the
anisotropic XY  spin chain in a magnetic field with Hamiltonian (see e.g.
\cite{Abanov})
\begin{equation}
H=-J\sum_{n=1}^N \left( h\sigma_n^x +\frac{1+\gamma}{2}\sigma_n^z
\sigma_{n+1}^z  +\frac{1-\gamma}{2}\sigma_n^y\sigma_{n+1}^y \right)  ~~,
\label{AnisotropicXY}
\end{equation}
is shown in Fig. \ref{AnisotropicXYFig} in terms of the topological 
charge $\tilde N_2$.  The lines $h=1$, $h=-1$ and ($\gamma=0$, $-1<h<1$),
which separate regions with different 
$\tilde N_2$, are lines of quantum phase transitions.

\subsubsection{Nullification of gap at quantum transition}

Because of the jump in $\tilde N_2$  [Fig. \ref{P-SpaceInstantonFig}({\bf
b})],  the transition cannot occur adiabatically. That is why the
energy gap must tend to zero at the transition, in the same way as it
occurs at the plateau-plateau transition in Fig.~\ref{FermionBosonFig}.
In the Ising model,  the energy spectrum 
$E^2(p)=g_x^2(p)+g_y^2(p)=4J^2\left((h -\cos(pa))^2 +
\sin^2(pa)\right)$ has a gap $E(0)=2J|h -1|$
which tends to zero at 
$h\rightarrow 1$ [Fig. \ref{P-SpaceInstantonFig} ({\bf b})]. However,
the nullification of the gap at the topological transition between the
fully gapped states with different topological charges is the general
property, which does not depend on the details of the underlying spin
system and is robust to interaction between the auxiliary fermions. 

The special case, when the gap does not vanish at the transition because
the momentum space is not compact, is discussed in Sec. 11.4 of
\cite{VolovikBook}.

\subsubsection{Dynamics of quantum phase transition and superposition of 
macroscopic states }

In the quantum Ising model of Eq.(\ref{Isings}) the ground state at 
$h<1$ represents the quantum superposition  of two ferromagnetic states
with opposite magnetization. However,  in the limit of infinite number of
spins  $N\rightarrow \infty$ this becomes  the Schr\"odinger's Cat -- the
superposition of two macroscopically different states. According to  Ref.
\cite{Immirzi} such superposition cannot be resolved by any measurements,
because in the limit $N\rightarrow \infty$  no observable has matrix
elements between the two ferromagnetic states, which are therefore
disjoint. In general, the disjoint states form the equivalence classes
emerging in the limit of infinite volume or infinite number of elements. 

Another property of the disjoint macroscopic states is that  their
superposition, even if it is the ground state of the Hamiltonian, can
never be achieved. For example, let us try to obtain the superposition of
the two ferromagnetic states at $h<1$ starting from the  paramagnetic
ground state at $h>1$ and slowly crossing the critical point $h=1$ of the
quantum phase transition. The dynamics of the time-dependent quantum
phase transition in this model has been discussed in Refs. 
\cite{Dziarmaga,ZurekQuantum}. It is characterized by the transition time 
$\tau_Q$ 
 which shows how fast the transition point is crossed: 
$1/\tau_Q= \dot h|_{h=1}$ . One may expect that if the transition occurs
adiabatically, i.e. in the limit $\tau_Q\rightarrow \infty$, the ground
state at $h>1$ transforms to the ground state at $h<1$. However, in the
limit $N\rightarrow \infty$ the adiabatic condition cannot be satisfied.
If $\tau_Q\rightarrow \infty$ but $\tau_Q\ll N^2/J$,   the transition
becomes non-adiabatic and the  level crossing
occurs with probability 1. Instead of the ground state at $h<1$  one
obtains the excited state, which represents two (or several)
ferromagnetic domains separated by the domain wall(s).  Thus in the
$N=\infty$ system  instead of the quantum superposition of the two
ferromagnetic states  the classical coexistence of the two ferromagnetic
states is realized. 

In the obtained excited state the translational and time reversal 
symmetries are broken.  This example of spontaneous symmetry breaking
occurring at $T=0$ demonstrates the general phenomenon that in the  limit
of the infinite system one can never reach the superposition of
macroscopically different states. On the connection between the process
of spontaneous symmetry breaking and the measurement process in quantum
mechanics see Ref. \cite{Grady} and references therein. Both processes
are emergent phenomena occurring in the limit of infinite volume $V$ of the
whole system. In finite systems the quantum mechanics is reversible.  For
general discussion of the symmetry breaking phase transition in terms 
of the disjoint limit Gibbs distributions emerging at $V\rightarrow\infty$ 
see the book by Sinai \cite{Sinai}. 

\section{Conclusion}

Here we discussed the quantum phase transitions which
occur between the vacuum states with the same symmetry
above and below the transition. 
Such a transition is essentially different from conventional
 phase transition   which is accompanied by the symmetry breaking.  The
discussed zero temperature phase transition is not the  termination point
of the line of the conventional 2-nd order phase transition: it is
either an isolated point
$(q_c,0)$ in the
$(q,T)$ plane, or the termination line of the 1-st order transition. This
transition is purely topological -- it is accompanied by the change of
the topology of fermionic Green's function in ${\bf p}$-space without
change in the vacuum symmetry.  The
${\bf p}$-space topology, in turn, depends on the symmetry of the
system.  The interplay between symmetry and   topology leads to variety
of vacuum states and thus to variety of emergent physical laws at low
energy, and to variety of possible quantum phase transitions. The more
interesting situations are expected for spatially inhomogeneous systems,
say for systems with topological defects in ${\bf r}$-space, where the
${\bf p}$-space topology, the ${\bf r}$-space topology, and symmetry are
combined (see Refs. \cite{Grinevich,Horava} and Chapter 23 in
\cite{VolovikBook}).

I thank Frans Klinkhamer for collaboration and Petr Horava
for discussions.
This work is supported in part
 by the Russian Ministry of Education and
Science, through the Leading Scientific School grant $\#$1157.2006.2, 
and by the European Science Foundation  COSLAB Program.


\begin{thebibliography}{99.}
\addcontentsline{toc}{section}{References}

%\bibitem{journ1} W. Frank, A. Seeger: Appl. Phys. A \textbf{3}, 66 (1988)

%\bibitem{proc1} W. Greiner, D.N. Poenaru: `Cluster Preformation in
%Closed- and Mid-shell Nuclei'. In:
%{\itshape Atomic and Nuclear Clusters, 2nd International Conference at Santorini, Greece,
%June 28--July 2, 1993}, ed. by G.S. Anagnostatos, W. von Oertzen (Springer, Heidelberg 1994)
%pp. 264--266

%\bibitem{fhjl1} F. Holzwarth, J. Lenz  et al.: {\itshape 1readme.
%Further Details on Layout and \LaTeX\, code}. (Springer, Berlin
%Heidelberg 1999)

%\bibitem{mono1} B. Jirgensons: {\itshape Optical Activity for Proteins
%and Other Macro-Molecules}, 2nd edn.
%(Springer, New York 1984)

%\bibitem{contr1} D.M. MacKay: `Visual Stability and Voluntary Eye
%Movements'. In: {\itshape Handbook of
%Sensory Physiology VII/3}. ed. by R. Jung (Springer, Berlin, Heidelberg 1973) pp. 307--331

%\bibitem{mono2} M. M\"uller, F.J. Becker: {\itshape On Generalized
%Hamiltonian Dynamics} (Cambridge University Press, Cambridge 1930)

%\bibitem{journ2} S. Nakamura, M. Senoh, N. Iwasa, S. Nagahama: Jpn. J.
%Appl. Phys. \textbf{34}, L797 (1995)
%W. Frank, A. Seeger: Appl. Phys. A \textbf{3}, 66 (1988)

%\bibitem{thesis1} D.W. Ross: Lysosomes and Storage Diseases. MA Thesis,
%Columbia University, New York (1977)




\bibitem{UnificationModel}
H. Georgi, S.L. Glashow:
%Unity of all elementary particle forces,
Phys.  Rev. Lett.\ \textbf{32}, 438 (1974)
%%CITATION = PRLTA,32,438;%%


\bibitem{Unification}
H. Georgi, H.R. Quinn, S. Weinberg:
%Hierarchy of interactions in unified gauge theories,
Phys. Rev. Lett. \textbf{33}, 451 (1974)
%%CITATION = PRLTA,33,451;%%

\bibitem{VolovikGorkov1985}
G.E. Volovik, L.P. Gorkov:
%Superconductivity classes in the heavy fermion systems,
Sov. Phys. JETP \textbf {61}, 843 (1985)

\bibitem{VollhardtWoelfle}
D. Vollhardt, P. W\"olfle:
{\itshape The Superfluid Phases of Helium 3}
(Taylor and Francis, London, 1990)

\bibitem{TopologyReview1} N.D. Mermin:  
%The topological theory of defects in ordered media,
Rev. Mod. Phys. \textbf {51}, 591 (1979)

\bibitem{VolovikBook} G.E. Volovik: {\itshape The Universe in a Helium
Droplet} (Clarendon Press,  Oxford, 2003)

\bibitem{Horava}  P. Horava:
%Stability of Fermi surfaces and $K$-theory,
Phys. Rev. Lett. \textbf{95}, 016405 (2005)

\bibitem{Khodel1990}  V.A. Khodel, V.R.  Shaginyan:
%Superfluidity in system with fermion condensate, 
JETP Lett. \textbf{51}, 553 (1990)

\bibitem{NewClass}  G.E. Volovik:
% A new class of normal Fermi liquids,   
JETP Lett. \textbf{53}, 222 (1991)

\bibitem{Shaginyan} V.R. Shaginyan, A.Z. Msezane, M.Ya. Amusia:
%Quasiparticles and order parameter near quantum phase
%transition in heavy fermion metals,  
Phys. Lett. A \textbf{338}, 393 (2005)

\bibitem{Khodel2005} V.A. Khodel, J.W. Clark, M.V. Zverev:
`Thermodynamic properties of Fermi systems with flat single-particle
spectra'
(cond-mat/0502292)

\bibitem{Khodel2005b}V.A. Khodel, M.V. Zverev, V.M.
Yakovenko:
%Curie law, entropy excess, and superconductivity in heavy
%fermion metals and other strongly interacting Fermi liquids, 
 Phys. Rev. Lett. \textbf{95}, 236402 (2005)


\bibitem{Volovik1994} G.E. Volovik:
%On Fermi condensate: near the saddle point and within  the
%vortex core, 
JETP Lett.  \textbf{59}, 830 (1994)



\bibitem{Sachdev} S. Sachdev: {\itshape Quantum Phase Transitions}
(Cambridge University Press,  Cambridge, 2003)

\bibitem{Lifshitz} I.M. Lifshitz: Sov. Phys. JETP  \textbf{11}, 1130 (1960);
I.M. Lifshitz, M.Y. Azbel, M.I. Kaganov: {\itshape Electron Theory of Metals}
(Consultant Press, New York, 1972)

\bibitem{Book1} G.E. Volovik: {\itshape Exotic Properties  of Superfluid
$^3$He}  (World Scientific, Singapore, 1992)

\bibitem{QPT} F.R. Klinkhamer, G.E. Volovik:
%Quantum phase transition for the BEC-BCS
%crossover in condensed matter physics and CPT
%violation in elementary particle physics, 
JETP Lett.  \textbf{80}, 343  (2004)

\bibitem{SplittingPreprint}   F.R. Klinkhamer, G.E. Volovik:  
%Emergent CPT violation from the splitting of Fermi points,
Int. J. Mod. Phys. A   \textbf{20}, 2795 (2005)
% hep-th/0403037.

\bibitem{Gurarie} V. Gurarie, L. Radzihovsky, A. V. Andreev: 
%Quantum phase transitions
%across $p$-wave Feshbach resonance, 
Phys. Rev. Lett.  \textbf{94}, 230403 (2005)

\bibitem{BotelhoPWavw} S.S. Botelho, C.A.R. Sa de Melo:
%Quantum phase transition in the BCS-to-BEC 
%evolution of $p$-wave Fermi gases, 
J. Low Temp. Phys.  \textbf{140}, 409 (2005)
% cond-mat/0504263.

\bibitem{Botelho} S.S. Botelho, C.A.R. Sa de Melo:
% Lifshitz transition in $d$-wave superconductors, 
Phys. Rev. B  \textbf{71}, 134507  (2005)


\bibitem{Borkowski} L.S. Borkowski, C.A.R. Sa de Melo: 
`From BCS to BEC superconductivity: 
Spectroscopic consequences' (cond-mat/9810370)

\bibitem{Duncan} R.D. Duncan, C.A.R. Sa de Melo:
%Thermodynamic properties in the evolution from BCS to Bose-Einstein 
%condensation for a $d$-wave superconductor at low temperatures, 
Phys. Rev.  B \textbf{62}, 9675 (2000)


\bibitem{WenZee} X.G. Wen, A. Zee:
%Gapless fermions and quantum order,
Phys. Rev. B  \textbf{66}, 235110 (2002)


\bibitem{Gubankova} E. Gubankova: 
`Conditions for existence of  neutral strange quark matter' (hep-ph/0507291)
 E. Gubankova, E. Mishchenko, F. Wilczek:
 % Gapless surfaces in anisotropic superfluids  cond-mat/0411238;
%Breached superfluidity via $p$-wave coupling,  
 Phys. Rev. Lett.  \textbf{94}, 110402  (2005)
K. Rajagopal, A. Schmitt: 
`Stressed pairing in conventional
color superconductors is unavoidable' (hep-ph/0512043) 
R. Casalbuoni: `Color Superconductivity in High Density QCD'
(hep-ph/0512198)
      

\bibitem{Pseudogap} N. Bergeal, J. Lesueur, M. Aprili, G. Faini, J. P.
Contour, B. Leridon:  
 `Direct test of pairing fluctuations in the pseudogap phase of
underdoped cuprates' 
 (cond-mat/0601265)  

\bibitem{Sadovskii}    M.V. Sadovskii: 
 `Models of the pseudogap state in 
high temperature superconductors' (cond-mat/0408489)

\bibitem{Sadovskii2} E.Z. Kuchinskii, M.V. Sadovskii:
`Non - Fermi liquid behavior in fluctuating gap model:
From pole to zero of the Green's function'
(cond-mat/0602406)  

\bibitem{Dzyaloshinskii}  I. Dzyaloshinskii: 
%Some consequences of the Luttinger theorem:   
%The Luttinger surfaces in non-Fermi liquids and Mott insulators
Phys. Rev.  Phys. Rev. B \textbf{68}, 085113 (2003)  

 



\bibitem{Phillips} T.D. Stanescu, P.W. Phillips, T.P. Choy:
cond-mat/0602280


\bibitem{Holstein} T. Holstein, R.E. Norton, P. Pincus: Phys. Rev. B
\textbf{8}, 2649 (1973)   

\bibitem{Reizer} M.Yu. Reizer: Phys. Rev. B \textbf{40},
 11571 (1989)  
M.Yu. Reizer: Phys. Rev. B \textbf{44},
 5476 (1991)  
W.E. Brown, J.T. Liu, H.C. Ren: Phys.Rev. D \textbf{62} 054013  (2000)
        
\bibitem{NonFLQCD}   A. Ipp, A. Gerhold, A. Rebhan: 
%Anomalous specific heat in
%high-density QED and QCD 
 Phys. Rev. D \textbf{69}, 011901(R) (2004)
T. Sch\"afer, K. Schwenzer:
%Non-Fermi liquid effects
%in QCD at high density, 
Phys. Rev. D \textbf{70}, 054007 (2004)
T. Sch\"afer:
`Fermionic Quasiparticles in QCD at High
Baryon Density' (hep-ph/0510044)


\bibitem{Khveshchenko} A.V. Chubukov, D.V. Khveshchenko: cond-mat/0604376

\bibitem{NeumannWigner} J. von Neumann, E. Wigner: Phys. Zeit. \textbf{30},
 467 (1929)     
   
\bibitem{StoneA} A.J. Stone: 
Proc. R. Soc. London A \textbf{351}, 141 (1976)

\bibitem{Arnold} V.I. Arnold: {\itshape Mathematical Method in Classical
Mechanics}  (Nauka, Moscow, 1979, in Russian) (Springer-Verlag, 1989)


\bibitem{Berry} M.V. Berry: 
Proc. R. Soc. London A \textbf{392}, 45 (1984)

\bibitem{Novikov} S.P. Novikov:
% Magnetic Bloch functions and vector bundles. 
%Typical dispersion laws and their quantum numbers. 
%Dokl. Akad. Nauk SSSR, \textbf{257}, 538 (1981)
Sov. Phys. Math. Dokl. \textbf{23}, 298 (1981)

\bibitem{VolovikDiabolic} G.E. Volovik: 
%Zeros in the fermion spectrum
%in superfluid systems as diabolical points,"
JETP Lett. \textbf{46}, 98 (1987) 

\bibitem{NielsenNinomiya} H.B. Nielsen, M. Ninomiya: 
%`Absence of neutrinos on a lattice. 
%I - Proof by homotopy theory', 
Nucl. Phys. B \textbf{185}, 20  (1981) 
%`Absence of neutrinos on a lattice. II -
%Intuitive homotopy proof',  
Nucl. Phys. B \textbf{193}, 173 (1981) 


\bibitem{WenPRL} X.G. Wen:
%Origin of gauge bosons from strong  quantum correlations
Phys. Rev. Lett.  \textbf{88}, 011602 (2002)

        
                                                                 
\bibitem{KlinkhamerCPT} F.R. Klinkhamer: 
`Lorentz-noninvariant neutrino oscillations: model and predictions' 
(hep-ph/0407200) 
 F.R. Klinkhamer:
 %Lorentz and CPT violation: a simple neutrino-oscillation model, 
 Nucl. Phys. B (Proc. Suppl.)  \textbf{149}, 209  (2005)
 % hep-ph/0502062.

 
\bibitem{Kajantie} K. Kajantie, M. Laine, K. Rummukainen,  M. 
Shaposhnikov: 
%Is there a hot electroweak phase transition at $m_HÊ \geq Êm_W$?
Phys. Rev. Lett. \textbf{77}, 2887 (1996)

 
\bibitem{Adler}  S.L. Adler: `Anomalies to all orders'. In:
  {\itshape Fifty Years of Yang-Mills Theory}, ed. by  G. 't Hooft   (World Scientific, 2006)
%hep-th/0405040
  
\bibitem{Blount} E.I. Blount:
%Symmetry properties of triplet  superconductors, 
Phys. Rev. B  \textbf{32}, 2935 (1985)
 
\bibitem{AnisotropyExperiment} H.J.H. Smilde, A.A. Golubov, Ariando,  G.
Rijnders, J.M. Dekkers, S. Harkema, D.H.A. Blank, H. Rogalla, H.
Hilgenkamp, 
%Admixtures to $d$-wave gap symmetry in untwinned 
%YBa$_2$Cu$_3$O$_7$ superconducting
%films measured by angle-resolved electron tunneling,  
Phys. Rev. Lett.  \textbf{95}, 257001 (2005)


\bibitem{Novoselov}  K.S. Novoselov, A.K. Geim, S.V. Morozov, D. Jiang,
M.I. Katsnelson, I.V. Grigorieva, S.V. Dubonos, A.A. Firsov:  
%Two-Dimensional Gas of Massless Dirac Fermions in Graphene
Nature \textbf{438},  197 (2005)

\bibitem{Gusynin}
S.G. Sharapov, V.P. Gusynin, H. Beck:
 Phys. Rev.  B  \textbf{69}, 075104 (2004)  
%Magnetic oscillations in planar systems with the Dirac-like 
%spectrum of quasiparticle excitations

 \bibitem{NovoselovQHE}  K. S. Novoselov, E. McCann, S. V. Morozov, V. I. Falko, M. I. Katsnelson, U. Zeitler, D. Jiang, F. Schedin, A. K. Geim:  `Unconventional quantum Hall effect and Berry's phase of 
 $2\pi$ in bilayer graphene' (cond-mat/0602565)

 \bibitem{Falko}  E. McCann, V.I. Fal'ko: 
% `Landau level degeneracy and quantum Hall effect 
%in a graphite bilayer'
%(cond-mat/0510237)   
 Phys. Rev. Lett. \textbf{96}, 086805 (2006)

\bibitem{Sato} M. Sato: `Nodal structure of
superconductors with time-reversal invariance and $Z_2$ topological
number' (cond-mat/0602445)
 
\bibitem{Kordyuk} A.A. Kordyuk, S.V. Borisenko, A.N. Yaresko,
S.-L. Drechsler, H. Rosner, T.K. Kim, A. Koitzsch, K.A. Nenkov, M.
Knupfer, J. Fink, R. Follath, H. Berger, B.
 Keimer, S. Ono,  Yoichi Ando:
%  Evidence for CuO conducting band
%splitting in the nodal direction of Bi$_2$Sr$_2$CaCu$_2$O$_{8+\delta}$,  
Phys. Rev. B  \textbf{70}, 214525 (2004)
S. V. Borisenko,  A. A. Kordyuk, V.
Zabolotnyy, J. Geck, D. Inosov, A. Koitzsch, J. Fink, M. Knupfer, B.
Buechner, V. Hinkov, C. T. Lin, B. Keimer, T. Wolf, S. G. Chiuzbaian, L.
Patthey, R. Follath:  
`Kinks, nodal bilayer splitting and interband scattering in YBCO' 
(cond-mat/0511596)
 

\bibitem{Ino} T. Yamasaki, K. Yamazaki,  A. Ino, M. Arita, H. Namatame,
M. Taniguchi, A. Fujimori, Z.-X. Shen, M. Ishikado, S. Uchida:
 `Unmasking the nodal quasiparticle dynamics in cuprate superconductors
using low-energy photoemission' (cond-mat/0603006)

  
 

  
 \bibitem{VolovikReentrant} 
 %Reentrant violation of special relativity in the low-energy corner
G.E. Volovik: JETP Lett.  \textbf{73}, 162 (2001)  
 
 \bibitem{KlinkhamerReentrant} F.R. Klinkhamer:  
`Possible new source of T and CP violation in neutrino oscillations'
(hep-ph/0601116) 

 \bibitem{Consoli}  M. Consoli:
%Spontaneous symmetry breaking and the $p\rightarrow 0$ limit
Phys. Rev. D \textbf{65}, 105017 (2002)
M. Consoli, E. Costanzo: Nuovo Cim. B  \textbf{119}, 393 (2004)    

                      



    

\bibitem{Volovik1993} G.E. Volovik: 
%Superconductivity with lines of gap nodes: 
%Density of states in the vortex
JETP Lett.  \textbf{58}, 469  (1993)

\bibitem{Balci} H. Balci, R.L. Greene:
% Anomalous change in  the field dependence of the 
%electronic specific heat of an electron-doped cuprate,
 Phys. Rev. Lett. \textbf{93}, 067001 (2004)  
 

\bibitem{Green}  W. Yu, B. Liang, R.L. Greene: 
%Magnetic-field dependence of the low-temperature
%specific heat of the
%electron-doped superconductor Pr1.85Ce0.15CuO4
Phys. Rev. B \textbf{72}, 212512 (2005) 




\bibitem{Ishikawa} K. Ishikawa, T. Matsuyama:
%Magnetic field induced multi component QED 
%in three-dimensions and quantum Hall effect,  
Z. Phys. C \textbf{33}, 41 (1986)
K. Ishikawa, T. Matsuyama:
%A microscopic theory of the quantum Hall effect, 
Nucl. Phys. B \textbf{280}, 523 (1987)

\bibitem{VolovikYakovenko}  G.E. Volovik, V.M. Yakovenko: 
%Fractional charge, spin and statistics of solitons in superfluid
%$^3$He film, 
J. Phys.: Condens. Matter  \textbf{1},  5263 (1989)



\bibitem{SQHE} G.E. Volovik:
`Fractional statistics and analogs
  of quantum Hall effect in superfluid $^3$He films'. In:  {\itshape Quantum Fluids and Solids - 1989} 
ed. by   G.G.Ihas, Y.Takano  
(AIP Conference  Proceedings , 1989)  \textbf{194},  pp. 136--146




\bibitem{MutualCS} Su-Peng Kou,  Xiao-Liang Qi, Zheng-Yu Weng:
%Spin Hall effect in a doped Mott insulator,
Phys. Rev. B \textbf{72}, 165114 (2005)


\bibitem{Zhitnitsky} D.T. Son, A.R. Zhitnitsky:
%Quantum anomalies in dense matter, 
Phys. Rev. D \textbf{70}, 074018 (2004)


\bibitem{Vafek} O. Vafek and A. Melikyan:
`Index theorem and quantum 
order of
$d$-wave superconductors in the vortex state' (cond-mat/0509258)

\bibitem{AngularMomentumPar} G.E. Volovik:
%Orbital momentum of vortices 
%and textures   due to spectral flow through the gap nodes:  Example of
%the $^3$He-A continuous vortex, 
JETP Lett. \textbf{61}, 958 (1995)

  
\bibitem{Rashba} E.I. Rashba: 
`Spin-orbit coupling and spin transport'
(cond-mat/0507007)
H.A. Engel, E.I. Rashba, B.I. Halperin: `Theory of Spin Hall Effects'
 (cond-mat/0603306)


\bibitem{Haldane} S.D.M. Haldane:
%Berry curvature on the Fermi surface: 
%Anomalous Hall effect as a topological Fermi-liquid property, 
Phys. Rev. Lett.  \textbf{93}, 206602 (2004)

\bibitem{EffectBandStructure}  G.E. Volovik:
%Poisson brackets scheme for vortex dynamics in superfluids 
%and superconductors and effect of band  structure of crystal, 
JETP Lett.  \textbf{64},  845  (1996)
%cond-mat/9610157.

\bibitem{JackiwRossi} R. Jackiw, P. Rossi: Nucl.Phys. B \textbf{190}, 681,
(1981)

\bibitem{WenStone} M. Stone: Ann. Phys.   \textbf{207}, 38 (1991)
X.G. Wen: Phys. Rev. B  \textbf{43}, 11025 (1991) 

\bibitem{Stonepxpy} M. Stone, R. Roy:
%Edge modes, edge currents, and gauge invariance in $p_x + ip_y$
%superfluids and superconductors
Phys. Rev. B \textbf{69}, 184511 (2004) 


\bibitem{Sengupta}  K. Sengupta, R. Roy, M. Maiti: `Spin-Hall effect in
triplet chiral superconductors and graphene' (cond-mat/0604217)

\bibitem{Dziarmaga} J. Dziarmaga:
%Dynamics of a quantum phase
%transition: exact solution of the quantum Ising model, 
Phys. Rev. Lett.  \textbf{95} 245701 (2005) 

\bibitem{Instanton} J.R. Hook, H.E. Hall:
%Orbital dynamics of $^3$He-A
%in the presence of a heat flow and a magnetic field, 
J. Phys. C  \textbf{12}, 783 (1979)
G.E. Volovik:
%Phase slippage without vortices and vector ${\bf l}$ oscillations in
%$^3$He-A,  
JETP Lett.  \textbf{27}, 573  (1978)

\bibitem{InstantonExp} D.N. Paulson, M. Krusius, J.C. Wheatley:
%Experiments on orbital dynamics in superfluid $^3$He-A,
Phys. Rev. Lett.  \textbf{36}, 1322  (1976)

\bibitem{Abanov} A.G. Abanov, F.  Franchini:
 %Emptiness formation 
%probability  for the anisotropic XY spin chain in a magnetic field,
Phys. Lett. A  \textbf{316},  342 (2003)
    
\bibitem{Immirzi} M. Fioroni,  G. Immirzi: 
 'How and why the wave
function collapses after a measurement'
 (gr-qc/9411044)



\bibitem{ZurekQuantum} W.H.  Zurek, U. Dorner, P. Zoller:
 %Dynamics of a quantum phase transition, 
 Phys. Rev. Lett.  \textbf{95}, 105701 (2005)

\bibitem{Grady}  M. Grady:
 `Spontaneous symmetry breaking as the mechanism  of quantum measurement' 
 (hep-th/9409049)


\bibitem{Sinai}  Ya.G. Sinai:
  {\itshape Theory of Phase Transitions,
International series in  natural philosophy}
(Pergamon Press, 1983)

\bibitem{Grinevich} P.G. Grinevich, G.E. Volovik:
%Topology of gap  nodes in superfluid $^3$He: 
%$\pi_4$ homotopy group for $^3$He-B  disclination, 
J. Low Temp. Phys.  \textbf{72}, 371  (1988)
M.M. Salomaa, G.E. Volovik: 
%Cosmiclike  domain walls in
% superfluid $^3$He-B: Instantons and diabolical points in (${\bf  k}$,
%${\bf r}$) space,
 Phys. Rev.  \textbf {37}, 9298  (1988) 
 M.M. Salomaa, G.E. Volovik: 
%Half-solitons in  superfluid $^3$He-A: 
 %Novel $\pi /2$-quanta of phase slippage, 
  J. Low Temp. Phys.  \textbf{74}, 319  (1989)



\end{thebibliography}
\end{document}